\documentclass[sigconf]{acmart}

\renewcommand\footnotetextcopyrightpermission[1]{} 

\usepackage{caption}
\usepackage{subcaption}
\usepackage{multirow}
\usepackage{url}
\usepackage{ragged2e}
\usepackage{afterpage}
\usepackage{tabularx}
\usepackage{paralist}
\usepackage{psfrag}
\usepackage{xpatch}
\usepackage{booktabs} 
\usepackage{balance} 
\usepackage{amsmath}
\usepackage{amsfonts}
\usepackage{setspace}
\usepackage{styles/citesort}
\usepackage[ruled,lined,linesnumbered]{algorithm2e}
\usepackage{balance}
\copyrightyear{2023}
\acmYear{2023}
\setcopyright{acmcopyright}\acmConference[ICCAD '24]{International Conference on Computer-Aided Design}{October 27--31, 2024}{New Jersey, USA}
\acmBooktitle{International Conference on Computer-Aided Design (ICCAD '24), October 27--31, 2024, New Jersey, USA}
\acmPrice{15.00}

\begin{document}

\title{Sky$^\varepsilon$-Tree: Embracing the Batch Updates of B$^{\varepsilon}$-trees through Access Port Parallelism on Skyrmion Racetrack Memory}

\author{Yu-Shiang Tsai}
\affiliation{%
	\institution{National Yang Ming Chiao Tung Uni.}
	\country{Taiwan}}
 \email{gacky1601@gmail.com}

\author{Shuo-Han Chen}
\affiliation{%
	\institution{National Yang Ming Chiao Tung Uni.}
	\country{Taiwan}}
\authornote{Corresponding authors}
\email{shch@nycu.edu.tw}

\author{Martijn Noorlander}
\affiliation{%
	\institution{University of Twente}
	\country{the Netherlands}}
 \email{m.s.noorlander@student.utwente.nl}

\author{Kuan-Hsun Chen}
\affiliation{%
	\institution{University of Twente}
	\country{the Netherlands}}
\email{k.h.chen@utwente.nl}
 
\renewcommand{\shortauthors}{Tsai et al.}

\begin{abstract}
Owing to the characteristics of high density and unlimited write cycles, skyrmion racetrack memory (SK-RM) has demonstrated great potential as either the next-generation main memory or the last-level cache of processors with non-volatility. Nevertheless, the distinct skyrmion manipulations, such as injecting and shifting, demand a fundamental change in widely-used memory structures to avoid excessive energy and performance overhead. For instance, while B$^{\varepsilon}$-trees yield an excellent query and insert performance trade-off between B-trees and Log-Structured Merge (LSM)-trees, the applicability of deploying B$^{\varepsilon}$-trees onto SK-RM receives much less attention. In addition, even though optimizing designs have been proposed for B$^{+}$-trees on SK-RM, those designs are not directly applicable to B$^{\varepsilon}$-trees owing to the batch update behaviors between tree nodes of B$^{\varepsilon}$-trees. Such an observation motivates us to propose the concept of Sky$^{\varepsilon}$-tree to effectively utilize the access port parallelism of SK-RM to embrace the excellent query and insert performance of B$^{\varepsilon}$-trees. Experimental results have shown promising improvements in access performance and energy conservation.
\end{abstract}

%

\thispagestyle{plain}
\pagestyle{plain}
\maketitle
\section{Introduction} \label{S:Introduction}
To meet the ever-growing memory demand of both computational and data-intensive applications, numerous nonvolatile random access memory (NVRAM) technologies~\cite{Chen:2010:DRC,pcm,reram,mram} have been proposed. Among these technologies, skyrmion racetrack memory (SK-RM)~\cite{Muhlbauer:2009:Science,Wang:2016} has emerged as a promising alternative to the existing SRAM-based last-level cache in processors~\cite{LLC} and the DRAM-based main memory in computer systems. SK-RM outperforms other NVRAMs with its high density, fast access speed, and DRAM-comparable write endurance~\cite{8107669}. Nevertheless, unlike conventional random access memory that can perform data updates at arbitrary locations, SK-RM relies on access ports to \textit{inject} or \textit{remove} skyrmions or non-skyrmions for representing data bits 1 and 0. Meanwhile, as the total number of access ports is smaller than that of data fields, \textit{shift} operations are required to align data fields with access ports. In other words, directly applying existing data layouts and data update policies on SK-RM could hinder the high-performance characteristics of SK-RM due to excessive insert and shift operations. Owing to the high density, non-volatility nature, and near-DRAM performance, SK-RM is suitable for hosting the indexing scheme, such as B-tree~\cite{btree} and B$^{\varepsilon}$-tree~\cite{betree}, of databases and file systems. Owing to the good query and upsert\footnote{An upsert is a term used to describe a combined operation of "update" and "insert" of key-value pairs in both B-trees and B$^{\varepsilon}$-trees.} performance of B$^{\varepsilon}$-tree, this study focuses on investigating the inefficiency of deploying B$^{\varepsilon}$-tree on SK-RM and facilitating data updates within B$^{\varepsilon}$-tree through the port parallelism of SK-RM. The technical difficulty of this study lies in \textit{how to layout B$^{\varepsilon}$-tree onto SK-RM for avoiding excessive insert and shift operations and leveraging the multi-port nature of SK-RM for performance enhancement.}

The origin of SK-RM dates back to 2008 when IBM first demonstrated the domain-wall racetrack memory (DWM)~\cite{Parkin:DWRM}. DWM stores data as magnetic domains on nanoscale magnetic wires, while domain walls refer to the boundaries between magnetic domains with different magnetic orientations. The key idea behind DWM is to use electric currents to move these domain walls along the magnetic wires, effectively reading and writing data. Data are accessed at the access ports that are integrated into the wires. The term ``racetrack'' comes from the similarity between the movement of magnetic domains and cars racing on the track. However, due to the size of each magnetic domain, DWM suffers from low cell density and high energy consumption. After a few years, SK-RM emerges and outperforms DWM with improved stability, higher cell density and lower energy consumption~\cite{Luo:2020:Nature}. Within SK-RM, data bit patterns are represented by the presence and absence of skyrmion, which is a nanoscale magnetic spin texture and resides on the racetrack. Similar to DWM, SK-RM also relies on shift MOS to move skyrmions through the access ports for both reading and writing. Owing to the great future prospective of SK-RM, researchers have investigated the word-based/bit-interleaved mapping architectures~\cite{10.1145/2333660.2333707,LLC}, readability enhancements~\cite{9417534}, access optimizations~\cite{9218642}, in-memory computation circuits~\cite{10.1145/3299874.3318015,electronics10020155}, the applicability of SK-RM to legacy algorithms (i.e., sorting)~\cite{9211559} and emerging applications (i.e., neural network computations)~\cite{skynn}. Nonetheless, \textit{the design issue of adopting SK-RM as the underlying memory devices for B$^{\varepsilon}$-tree receive much less attention.}

By investigating the I/O behavior of B-trees, B$^{\varepsilon}$-trees were proposed by Brodal and Fagerberg~\cite{betree} to enhance the update performance by orders of magnitude faster than B-trees and still retain identical query performance. Many researchers foresee B$^{\varepsilon}$-trees as a sweet spot on the trade-off curve of update-query performance, thus facilitating ongoing research and file system designs based on B$^{\varepsilon}$-trees~\cite{DBLP:journals/usenix-login/BenderFJJKPY015, 188458}. The main concept of B$^{\varepsilon}$-trees is to include a buffer at the internal tree nodes. Then, unlike B-trees requiring flush individual upserted key-value pairs down to the leaf node, B$^{\varepsilon}$-trees first buffer key-value pairs at the root nodes and flush those buffered key-value pairs to the next level of the tree in a recursive fashion after the buffer is full or based on certain conditions. In this study, such flushes are referred to as \textit{batched update}. Based on SK-RM, the design to adopt different variations of B-trees has been discussed. For instance, Chang et al. propose Sky-tree, which is an SK-RM-friendly B$^{+}$-tree, to enable bit-level binary search and intratrack node splitting for enhancing the performance of B$^{+}$-trees on SK-RM~\cite{9925692}. However, the  B$^{+}$-tree-based design cannot be applied to B$^{\varepsilon}$-trees directly for the following reasons. First, unlike the batched update behavior of B$^{\varepsilon}$-trees, B$^{+}$-trees still flush upserted key-value pairs individually. In other words, previous designs have not considered the batch update behavior. Secondly, even though B$^{+}$-trees extend the original B-tree to utilize a double-linked list to maintain its leaf node for easy in-order traversal, data (i.e., the value of key-value pairs) can only be retrieved at leaf nodes. Differently, the buffer of B$^{\varepsilon}$-trees' nodes allows data to be retrieved at internal nodes. Thus, the data retrieval process should also be reconsidered. In summary, \textit{aforementioned differences call for an urgent need to enable an efficient B$^{\varepsilon}$-trees on SK-RM.}

Deploying B$^{\varepsilon}$-trees directly onto SK-RM, treating SK-RM like conventional RAM, allows the features of B$^{\varepsilon}$-trees to function but may cause unnecessary overhead, including excessive insert and shift operations. In particular, the number of shift operations can be diminished if each of the key-value pairs can be updated paralleled during batched updates. On the other hand, as multiple buffers are included at the internal nodes of B$^{\varepsilon}$-trees, a key-value pair is written multiple time when compared with both the B-trees and B$^{+}$-trees. In other words, the lengthy and power-consuming skyrmion injections are performed multiple times, which becomes an unwanted performance overhead. 

\noindent\textbf{Our Contributions:} To address the inefficiency issues that are encountered during deploying B$^{\varepsilon}$-trees on SK-RM, this study proposes an innovative concept of \textit{Sky$^{\varepsilon}$-tree}, which consists of the virtual buffer encoding and parallel port updates. In a nutshell, the contributions of this study are summarized as follows:
\begin{compactitem}
	\item This study identifies the key difference of B$^{\varepsilon}$-trees and proposes leveraging the access port parallelism of SK-RM to enhance the batched updates of B$^{\varepsilon}$-trees.
	\item To resolve the issue of excessive skyrmion injections during flushing key-value pairs recursively within B$^{\varepsilon}$-trees, the proposed Sky$^{\varepsilon}$-tree introduces the virtual buffer encoding to separate keys and values.
	\item The proposed Sky$^{\varepsilon}$-tree is examined under both the \textit{word-based} and \textit{bit-interleaved mapping} architectures, based on an in-house simulator and enhanced RTSim~\cite{khan2019rtsim}, respectively.
	\item Experimental results show that the Sky$^{\varepsilon}$-tree outperforms the baseline method by up to 77.22\% and 80.49\% under the word-based and bit-interleaved mapping architectures.
\end{compactitem}

The rest of this paper is organized as follows. The background of SK-RM, B$^{\varepsilon}$-trees and research motivation are summarized in Section~\ref{S:background}. Then, Section~\ref{S:sky} describes the proposed Sky$^{\varepsilon}$-tree. Then, a series of experiments are conducted in Section~\ref{S:evaluation} for evaluation. Finally, Section~\ref{S:conclusion} concludes this paper with research remarks.

\section{Background and Motivation} \label{S:background}
We start with the background of SK-RM, including its mapping architectures and its distinct operations (in Section~\ref{sub:SK-RM}). Afterward, various tree-based index schemes, such as B-trees and B$^{\varepsilon}$-trees, are introduced (in Section~\ref{sub:index}). Finally, a motivational example is provided to reveal the inefficiency of conducting the batch update of B$^{\varepsilon}$-trees directly on SK-RM (in Section~\ref{sub:moti}).

\subsection{Skyrmion Racetrack Memory}\label{sub:SK-RM}
In skyrmion racetrack memory (SK-RM), nanoscale particle-like spin-swirling configurations known as magnetic skyrmions are used to represent data bits on a magnetic nano track, also known as a racetrack, that is further divided into hundreds of tiny magnetic regions, with each representing an information (bit) carrier. Data is accessed through access ports, typically consisting of a magnetic tunneling junction (MTJ) along with an access transistor. 
For current SK-RM prototypes, a possible size of a nanotrack is 64 bits, and a possible interport separation or data segment is 8 bits. SK-RM has been considered as the medium of different layers, such as scratchpad memory, persistent storage devices, and main memory. This work assumes the target system is equipped with SK-RM as main memory, where the data structure of B$^{\varepsilon}$-trees is mapped to.

SK-RM has four basic operations, namely \texttt{shift}, \texttt{detect}, \texttt{remove}, and \texttt{inject}. Given the fixed positions of access ports, the \texttt{shift} operation is to move the nanotrack to align with access ports before access. The \texttt{detect} operation is to examine the presence of skyrmion elements on an aligned region. The presence of a skyrmion element represents bit 1; otherwise, the bit is said to be 0. The \texttt{remove} operation applies a remove current to shift out the skyrmion on the aligned region (i.e., removing a bit 1); by contrast, to squeeze out a bit 0, only a shift current is needed like a \texttt{shift} operation. On the other hand, 
When the data bit value is 1, the injector creates a skyrmion and the \textit{inject} current is supplied to the bottom of the access ports to move the skyrmion into the intersecting bit zone. If the data bit value is 0, no injection is necessary because the lack of skyrmions corresponds to data bit 0, known as a non-skyrmion state. Note that after both the \texttt{remove} operation and the \texttt{inject} operation, the nanotrack is also shifted subsequently. 

\begin{figure}[t]
	\centering
	\includegraphics [width=3.in]{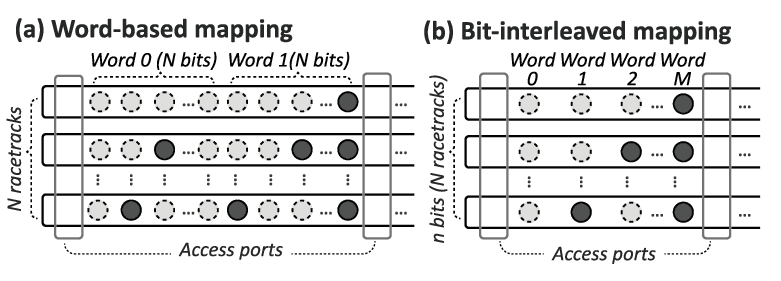}
	\vspace{-0.1in}
	\caption{Word-based and bit-interleaved mapping.}
	\label{fig:mappingExample}
	\vspace{-0.05in}
\end{figure}

\begin{figure}[t]
	\centering
	\vspace{-0.05in}
	\includegraphics[width=3.4in]{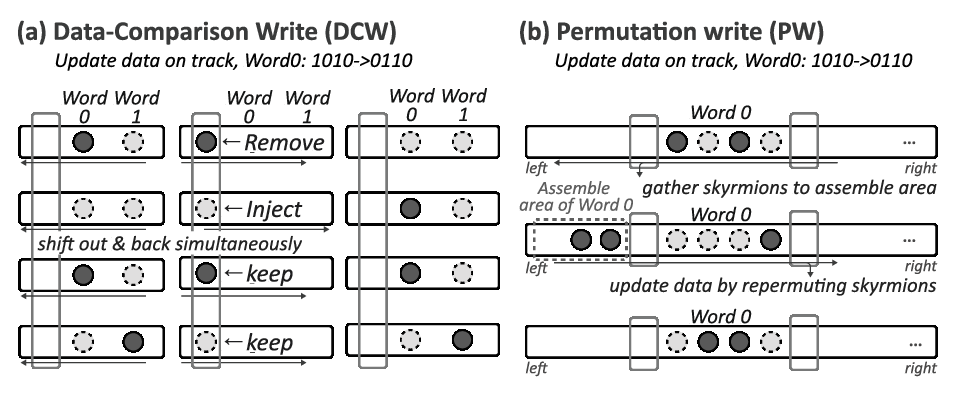}
	\vspace{-0.2in}
	\caption{Bit-comparison and permutation-write strategies.}
	\label{fig:writingExample}
	\vspace{-0.1in}
\end{figure}

Various data layouts have been proposed for racetrack memories, such as the word-based manner~\cite{9218642, 9211559} and the bit-interleaved manner~\cite{10.1145/2333660.2333707, 9211559}. As shown in Figure~\ref{fig:mappingExample}, with the word-based manner, data bits are encoded and stored consecutively on the nanotrack(s), whereas data bits are alternatively allocated to different cells in a round-robin fashion with the bit-interleaved manner. In this work, both organizations are considered during the evaluation. For writing strategies on SK-RM, several approaches have been proposed, e.g., data-comparison (DCW) and permutation-write (PW), as shown in Figure~\ref{fig:writingExample}. When compared with the na\"ive strategy that removes all existing skyrmions on the track (i.e., \texttt{shift} and \texttt{remove}) via the access port, both DCW and PW are more advanced writing strategies that can preserve existing skyrmions for composing future data patterns~\cite{9218642, LLC}. Based on the bit-interleaved mapping, DCW first performs \texttt{detect} operations to identify pattern differences. Then, \texttt{remove} and \texttt{inject} operations are utilized to flip data bits accordingly. On the other hand, based on the word-based mapping, PW re-permutes the existing Skyrmions in the old data to form the newly written data. These advanced strategies can effectively reduce the time- and energy-consuming \texttt{inject} operations for better performance and energy efficiency. In this work, we show that PW is not applicable for B$^{\varepsilon}$-tree optimization, but its ability to reuse existing skyrmions can still be retained by the Sky$^{\varepsilon}$-tree.

\vspace{-0.1in}
\subsection{Tree-based Indexing Schemes and B$\varepsilon$-Tree}\label{sub:index}
For managing indexing keys for database and file systems, as a general class, $B$-tree has many variances, such as $B^+$-tree and Log-Structured Merge (LSM)-trees. The main idea behind this class of designs is to leverage the structure of self-balancing search tree to make operations over large data efficiently, i.e., in a logarithmic time complexity for inserts, updates, and deletes. With different merits, both $B^+$-trees and LSM-trees have their place for data storage and retrieval~\cite{DBLP:conf/fast/QiaoCZLL022}, whilst $B^\varepsilon$-tree yields both high query and insert performance~\cite{DBLP:journals/usenix-login/BenderFJJKPY015, betree}.
As one of the most famous variations, $B^+$-tree is well-known for its high query performance, and it empowers almost all the rational database management systems. In the $B^+$-tree structure, all values are maintained in leaves, while keys are stored in interior nodes. Moreover, the leaves of the $B^+$-tree are linked to provide ordered access as well.
Compared with $B$-tree, the height of $B^+$-tree is greatly reduced and thus the number of storage accesses needed to find the target key. However, it has a low storage space usage efficiency and fits poorly to write-intensive workloads (i.e., updating the tree requires multiple random writes).

As a contender to $B^+$-tree, the LSM-tree is known for its advantages in terms of space usage and write amplification. 
In an LSM-tree, incoming data is batched together in a temporary sorting area, and once it fills up, they are flushed to storage as a sequential log (termed SSTable). These SSTables are organized as a hierarchy of levels, and the amount of data in each level increases as the levels get higher.
By merging SSTables together, LSM-trees remove duplicates and outdated entries, optimizing space usage and reducing the overhead of indexing updates. Despite its advantages, however, LSM-trees may suffer from the lower read performance (e.g. unproductive searches within queries) and compaction overhead.

Similar to $B^+$-tree, $B^\varepsilon$-tree also stores key-value pairs in the leaves and pivots for navigation towards leaves in internal nodes. Besides storing pivots, nevertheless, the concept of buffers allocated in internal nodes makes $B^\varepsilon$-tree outstanding. Newly inserted key-value pairs are buffered in the root node as "messages" and flushed down to the next level only when the buffer is full. As these messages are moved together (so-called batched update), it results in fewer I/O costs than $B$-trees flushing individual pairs. The value of $\varepsilon$ is a tuning parameter between 0 and 1 that determines how much space in each internal node is reserved for pivots and messages. Note that when $\varepsilon = 1$, $B^\varepsilon$-tree approximates a regular $B$-tree.

\subsection{Motivational Examples}\label{sub:moti}
\begin{figure}[h]
    \centering    
    \includegraphics [width=1.8in]{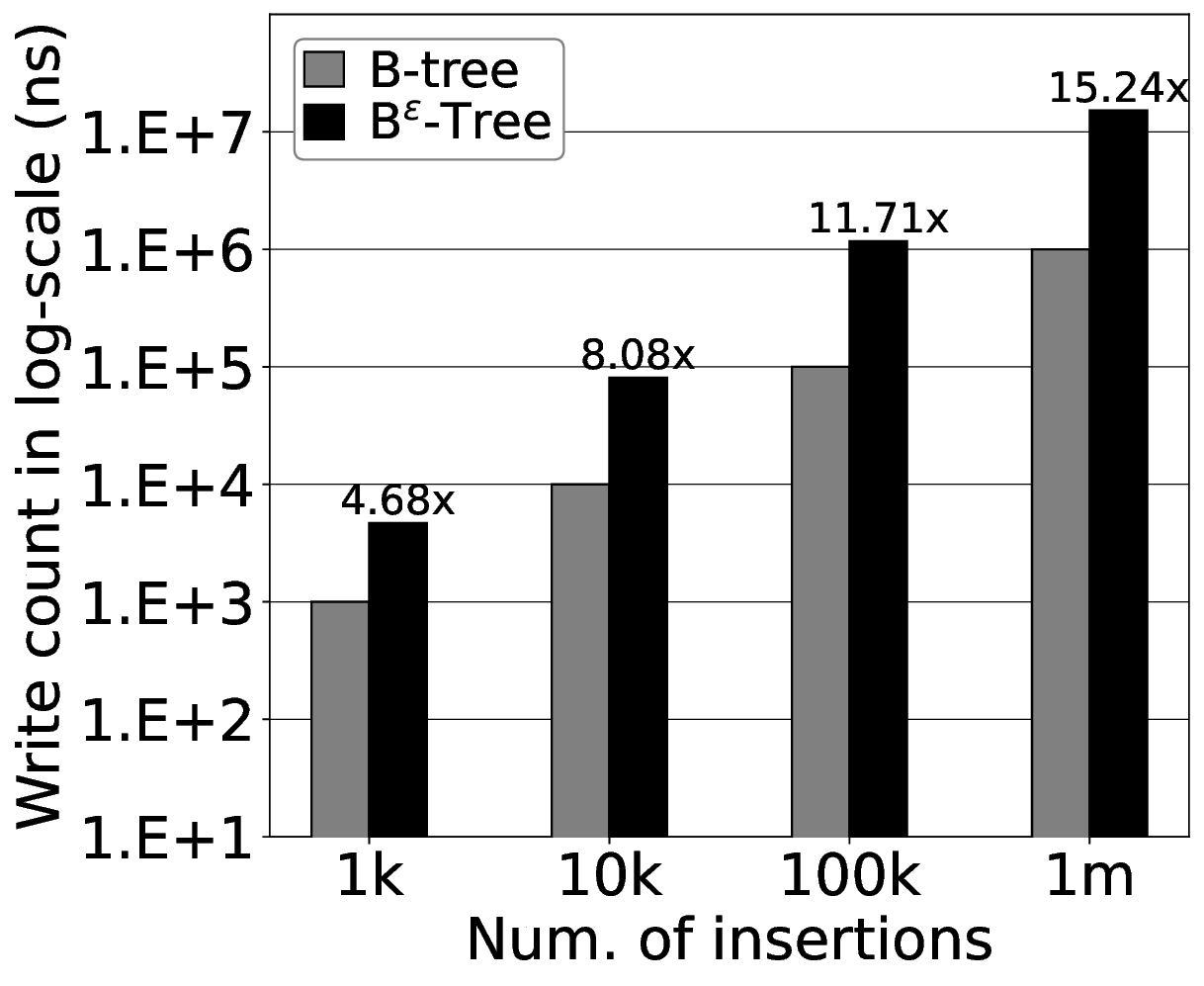}
    \vspace{-0.1in}
    \caption{Comparison of key-value pairs' write counts with different numbers of key-value pairs insertions.}
    \label{fig:copies}
    \vspace{-0.05in}
\end{figure}

The motivation of this work is in twofold. First, we observe that, in B$^\varepsilon$-trees, inserted key-value pairs could be rewritten at internal nodes multiple times before reaching the leaves, as shown in Figure~\ref{fig:copies}. In this figure, an experiment is conducted to compare the write counts of key-value pairs under B-trees and B$^\varepsilon$-trees with different numbers of insertions. According to the result, the difference grows up to $15.24$ times with $1$ million key-value pair insertions. Since such a rewritten behavior involves the SK-RM specific operations under both word-based and bit-interleaved mapping with BCW and PW strategies, the rewritten behavior induces high energy and latency on SK-RM, differing from the conventional setup where DRAM is used as the main memory. Thus, an SK-RM-friendly mechanism is needed to reduce such costly writing overhead.

Second, the state-of-the-art optimization strategies for SK-RM, such as permutation write~\cite{9218642} and Sky-tree~\cite{9925692}, focus on the overhead of individual write by reusing existing skyrmions and reducing excessive shifting operations. However, the batch update behavior has not been considered. For instance, when writing multiple key-value pairs on the same racetrack, it is possible to utilize a set of shift operations and conduct operations on each port in parallel to reduce overhead. Hence, we argue that a specific optimization strategy for $B^\varepsilon$-trees can be devised to embrace the batched update behavior and avoid writing each key-value pair individually.

\section{Sky$^{\varepsilon}$-tree} \label{S:sky}
In this section, the concept and detailed designs of Sky$^\varepsilon$-tree are introduced. After presenting the data layout under the word-based mapping architecture is illustrated (in Section~\ref{sub:overview}), we introduce two important concepts, namely the virtual buffer encoding (in Section~\ref{sub:VBE}) and the parallel port update (in Section~\ref{sub:PPU}). Finally, we discuss how to deploy Sky$^\varepsilon$-tree on the bit-interleaved mapping architecture (in Section~\ref{sub:bim}).

\subsection{Word-based mapping for the Sky$^{\varepsilon}$-tree}\label{sub:overview}
\begin{figure}[h]
	\centering
        \vspace{-0.05in}
        \includegraphics [width=3.1in]{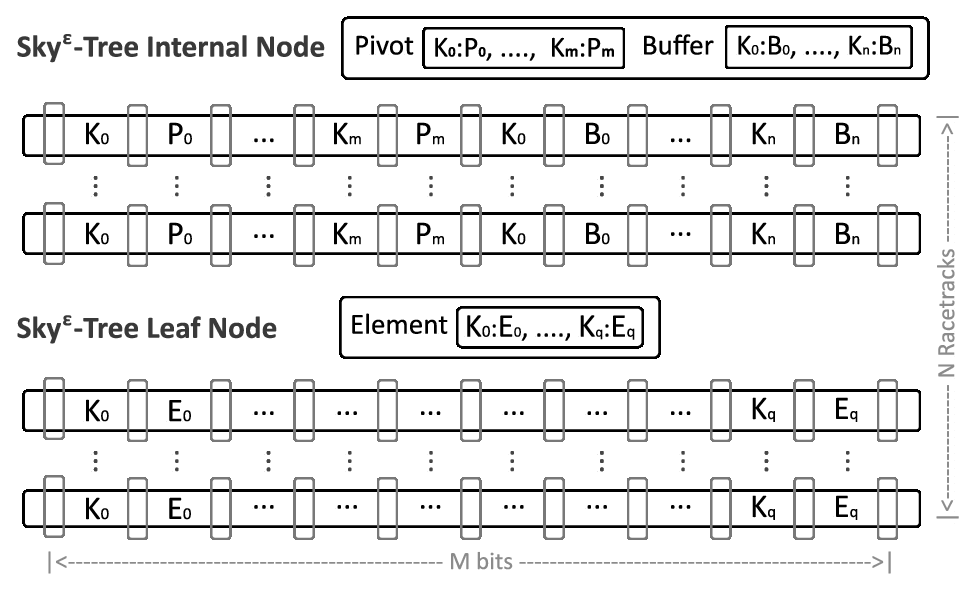}
        \vspace{-0.1in}
        \caption{Word-based data layout of the Sky$^{\varepsilon}$-tree, in which an internal node occupies a racetrack for its \textit{pivots} and \textit{buffer}, and a leaf node stores its \textit{elements} on the same track. K, P, B, and E denote the keys, pivots, buffer, and elements.}
        \label{fig:word}
        \vspace{-0.05in}
\end{figure}

To facilitate the data processing of B$^{\varepsilon}$-trees on SK-RM, this study proposes the Sky$^{\varepsilon}$-tree to delicately configure the data layout of B$^{\varepsilon}$-trees and redesign the data update methods with the awareness of SK-RM characteristics. The design principles of the Sky$^{\varepsilon}$-tree can be summarized as follows. First, the proposed Sky$^{\varepsilon}$-tree leverages the multiple access ports nature of SK-RM to perform the batched updates of B$^{\varepsilon}$-trees in parallel. Second, to avoid excessive skyrmion insertions, the key and value fields are separated when data are in the buffer of internal nodes. Then, keys and values are rejoined when reaching leaf nodes. Figure~\ref{fig:word} shows that the Sky$^{\varepsilon}$-tree follows the original design of B$^{\varepsilon}$-trees to include \textit{pivots} and \textit{buffer} for internal tree nodes or \textit{elements} for lead nodes. In other words, instead of altering the original data structure of B$^{\varepsilon}$-trees, the proposed Sky$^{\varepsilon}$-tree confines its alterations to data layout and update methods without affecting the original functionalities of B$^{\varepsilon}$-trees.

In the implementation of this paper, \textit{pivots}, \textit{buffer}, and \textit{elements} are key-value pairs. Pivots store the key value and the corresponding pointer to child nodes, while buffer temporarily holds those upserted key-value pairs from the users of the Sky$^{\varepsilon}$-tree. Then, those inserted key-value pairs are eventually stored as \textit{elements} at leaf nodes. Currently, an interport distance stores either a key or a value, and one access port is assigned to that. However, the number of access ports can be lowered to increase the storage density of SK-RM or increased for higher port parallelism. Accordingly, the virtual buffer encoding and parallel port updates are introduced based on the data layout shown in Figure~\ref{fig:word}. Notably, this figure shows the data layout of Sky$^{\varepsilon}$-tree under the word-based mapping architecture. The applicability of deploying the proposed Sky$^{\varepsilon}$-tree to bit-interleaved mapping architecture is discussed in Section~\ref{sub:bim}.

\subsection{Virtual Buffer Encoding}\label{sub:VBE}
\begin{figure}[h]
	\centering
	\vspace{-0.05in}
	\includegraphics [width=3.3in]{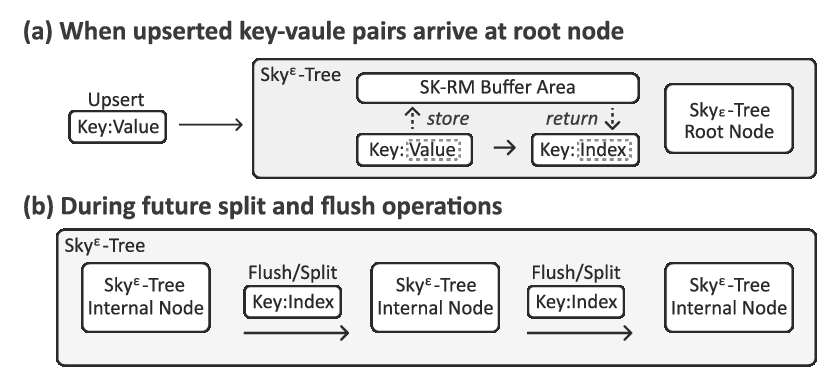}
	\vspace{-0.1in}
	\caption{Working principle of the virtual buffer encoding, in which values of upserted key-value pairs are separated and replaced by indices with less number of bits for reducing skyrmion injections during flush/split operations.}
	\label{fig:VBE}
	\vspace{-0.05in}
\end{figure}

Even though B$^{\varepsilon}$-trees eliminate individual updates for upserted key-value pairs and perform batched updates, inserted key-value pairs could be rewritten at internal nodes multiple times before reaching the leaf node. Conventionally, such behavior leads to good insertion performance when DRAM and secondary storage (i.e., HDDs or SSDs) are utilized to store B$^{\varepsilon}$-trees. This is because, in terms of latency and energy consumption, DRAM has similar read and write costs. Nevertheless, writing data on SK-RM induces higher latency than reading (See Table~\ref{skyr_laten_opera}), and distinct shift operations are required during every read and write. Writing upserted key-value pairs multiple times could aggravate the overall performance. To resolve the aforementioned concerns, the proposed Sky$^{\varepsilon}$-tree separates keys and values of those upserted key-value pairs from users. Meanwhile, the value field in each buffer entry in the internal nodes is replaced with indices with a smaller number of bits to point to the reserved buffer area for those values. Figure~\ref{fig:VBE} shows the working principle of the proposed virtual buffer encoding.

As shown in Figure~\ref{fig:VBE}(a), when an upserted key-value pair arrives at the tree root, the key and value are first separated. A buffer entry is then allocated from the buffer area with a returned index. Later, the key and the returned index are combined and inserted into the buffer of the root node. Then, as Figure~\ref{fig:VBE}(b) shows, during future flush and split operations of the Sky$^{\varepsilon}$-tree, the number of rewritten bits between the buffer of each internal node can be lowered to reduce the number of skyrmion injections. To understand the memory overhead of the indices, the experimental results are referred to as an example. According to the experimental results, the average required bits of the index field are 8 and 17 bits for datasets with the size of 1 thousand and 1 million key-value pairs. Based on the number of bits, the storage overhead of maintaining indices of the buffer area can be calculated by multiplying the number of required bits and the number of maximum entries. Under the worst-case scenario, when the buffer area is full, the overhead of maintaining the indices in the buffer of internal nodes is 256 and 278528 bytes for 1 thousand and 1 million datasets. If each key-value pair in datasets is 16 bytes, the overhead is 1.6\% and 1.74\%, respectively, when compared with the size of datasets. Compared with the reduction of skyrmion injections, this additional overhead is relatively acceptable, as shown in the experimental results (in Section~\ref{S:evaluation}).

\subsection{Parallel Port Updates}\label{sub:PPU}
\begin{figure}[h]
	\centering
	\vspace{-0.05in}
	\includegraphics [width=3.3in]{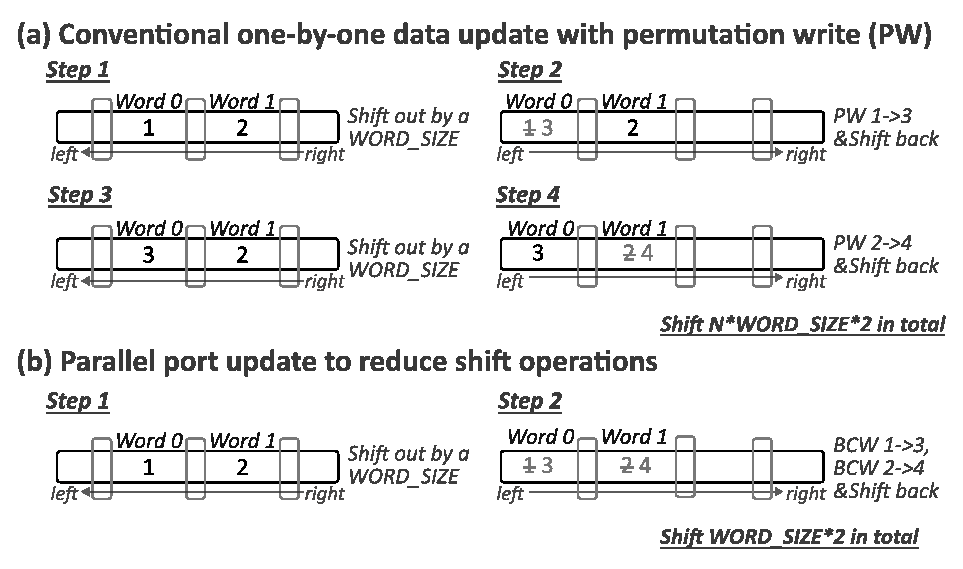}
	\vspace{-0.1in}
	\caption{Comparison of the conventional one-by-one update and the proposed parallel port update, which shows that the parallel port update can effectively reduces the number of required shift operations.}
	\label{fig:p}
	\vspace{-0.05in}
\end{figure}

In the design of B$^{\varepsilon}$-trees, the main improvement of upsert performance comes from the batched update behavior, in which multiple key-value pairs are buffered within internal nodes and flushed down the tree together. However, when deploying B$^{\varepsilon}$-trees on SK-RM, the batched update behavior could cause excessive shift operations and should be reconsidered to avoid excessive shift operations. For instance, if every key-value pair is written individually during flush operations, Sky-RM induces $N \cdot WORD\_SIZE \cdot 2$ shifts to write $N$ key-value pairs in the same batch. The $WORD\_SIZE$ denotes the size of a key or a value, and multiplying by $2$ is because data bits are shifted out and back during SK-RM data updates. In this study, the size of a key and a value are assumed to be identical. The above behavior can be illustrated as Figure~\ref{fig:p}(a).

In Sky$^{\varepsilon}$-tree, instead of writing key-value pairs of the same batch individually, the data writes are conducted in parallel by utilizing the multiple access ports nature of SK-RM. In other words, this study points out that, as shift operations move all skyrmions on the same racetracks, key-value pairs of the same batch update should share a single set of shift operations. Accordingly, the method of parallel port updates is included to leverage the access port parallelism of SK-RM to perform batched write and perform a single set of shift operations to move skyrmions out and back across access ports. The total number of shift operations is reduced to $WORD\_SIZE \cdot 2$. Figure~\ref{fig:p}(b) summarizes the method of parallel port updates.

\begin{figure}[h]
	\centering
	\vspace{-0.05in}
	\includegraphics [width=3.3in]{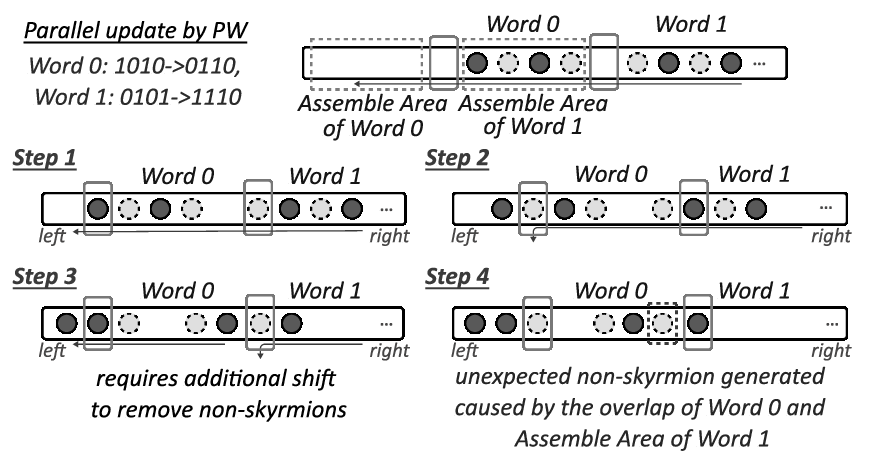}
	\vspace{-0.1in}
	\caption{Issue of deploying PW within the proposed parallel port update, which shows that additional shift are required to remove non-skyrmions. Meanwhile, unexpected non-skyrmions are emerged in Word 1 while composing the next data pattern for Word 0.}
	\label{fig:pwissue}
	\vspace{-0.05in}
\end{figure}


As the state-of-the-art writing strategy, permutation-write (PW) however is not directly applicable. Firstly it  
requires a buffer region for a word to recompose its skyrmions for representing new data patterns, updating multiple words in parallel on a single racetrack occupies the buffer region of subsequent words. In addition, when moving all skyrmions on the racetrack simultaneously, it becomes infeasible for PW to remove non-skyrmions when shifting skyrmions into the buffer region, as shown in Figure~\ref{fig:pwissue}.  
To retain the ability to reuse existing skyrmions, the proposed parallel port updates perform bit-comparison write (BCW) at every access port in parallel while updating data on SK-RM. 
The above operations is summarized as Algorithm~\ref{algo} for a better understanding.

\begin{algorithm}[t]
    \setstretch{1}
    \DontPrintSemicolon
    \caption{Bit-Comparison Write (BCW)}\label{algo}
    \KwIn {$P$: the previous data pattern of words in the batch}
    \KwIn {$N$: the to-be-written data pattern of words in the batch}
    \KwIn {$word\_size$: the number of bits per word}
    \KwIn {$batch\_size$: the number of words in the batch}
    \tcp {Shift out the racetrack}
    \For{$i = 0$ \textbf{to} $word\_size-1$} {
         \textbf{Shift} right by 1 bit\; 
    }
    \tcp {Update multiple words in parallel while shifting back}
    \For{$i = 0$ \textbf{to} $word\_size-1$}{
        \For{$j =0$ \textbf{to} $batch\_size-1$}{
            $old\_bit \leftarrow$ \textbf{Detect} the value of i$^{th}$ bit in $P[j]$\;
            $new\_bit \leftarrow$ \textbf{Detect} the value of i$^{th}$ bit in $N[j]$\;
            \If{$old\_bit == 0$  and $new\_bit == 1$ }{
                 \textbf{Inject} a new Skyrmion\;
            }
            \ElseIf{$old\_bit == 1$  and  $new\_bit == 0$}{
                \textbf{Remove} an old Skyrmion\;
            }
            \Else{
                \textbf{Keep} the Skyrmion or non-Skyrmion in place\;
            }
        }
        \textbf{Shift} left by 1 bit\; 
    }
\end{algorithm}

\subsection{Adopting Bit-interleaved Mapping}\label{sub:bim}
\begin{figure}[h]
	\centering
	\vspace{-0.05in}
	\includegraphics [width=2.9in]{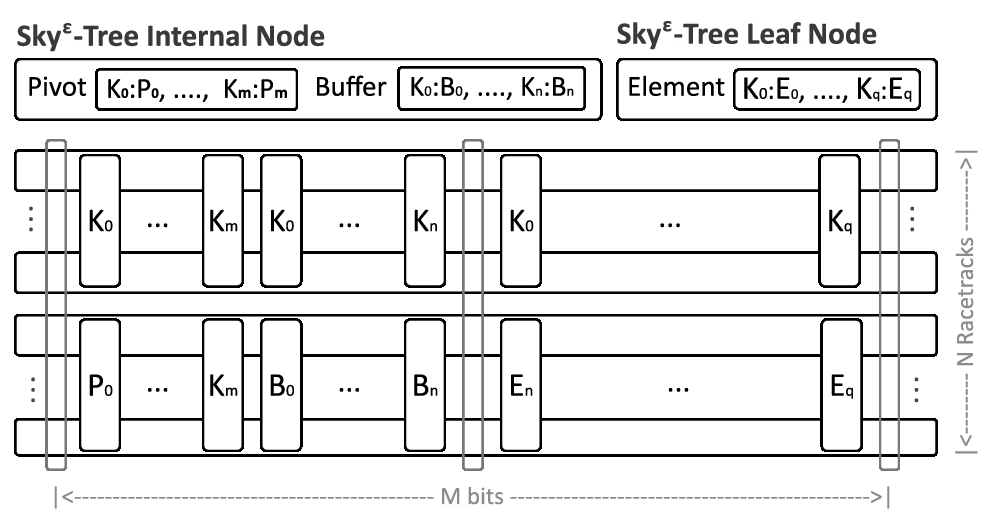}
	\vspace{-0.1in}
	\caption{Bit-interleaved data layout of the Sky$^{\varepsilon}$-tree, in which key-value pairs of pivots, buffer, and elements are mapped vertically across multiple racetracks. K, P, B,
and E denote the keys, pivots, buffer, and elements.}
	\label{fig:bit-inter}
	\vspace{-0.05in}
\end{figure}

As the bit-interleaved mapping is another distinct mapping architecture for SK-RM, featuring different update characteristics, we also investigate the possibility of adopting this mapping architecture for the proposed Sky$^{\varepsilon}$-tree. First, to realize the port parallelism property under bit-interleaved mapping, the data layout of Sky$^{\varepsilon}$-tree is configured as Figure~\ref{fig:bit-inter}. For key-value pairs in the pivots, the buffer, and the element region of internal and leaf nodes, all key-value pairs are layout vertically onto the SK-RM and the number of required racetracks equals the $WORD\_SIZE \cdot 2$ since one word stores the key and the other word holds the value. Notably, unlike the word-based Sky$^{\varepsilon}$-tree stores the key-value pairs of a node sequentially, the bit-interleaved Sky$^{\varepsilon}$-tree groups a fixed number of nodes and gathers the key-value pairs at the same offset of each node in an interport region. Afterwards, all key-value pairs of the same node can be aligned to access ports with the same set of shift operations. Identical to word-based Sky$^{\varepsilon}$-tree (See Figure~\ref{fig:word}), the interport distance equals to the $WORD\_SIZE$ to maintain the same density of access ports. Second, the virtual buffer encoding is still effective, as the primary goal is to use fewer bits for representing the values of upserted key-value pairs. Therefore, no additional adaption is required. In summary, \textit{the proposed Sky$^{\varepsilon}$-tree can be deployed on both word-based and bit-interleaved mapping architectures while benefiting from both proposed methods.}

\begin{figure*}
	\centering
	\begin{minipage}[h]{0.36\textwidth}
		\centering
		\includegraphics[height=1.65in] {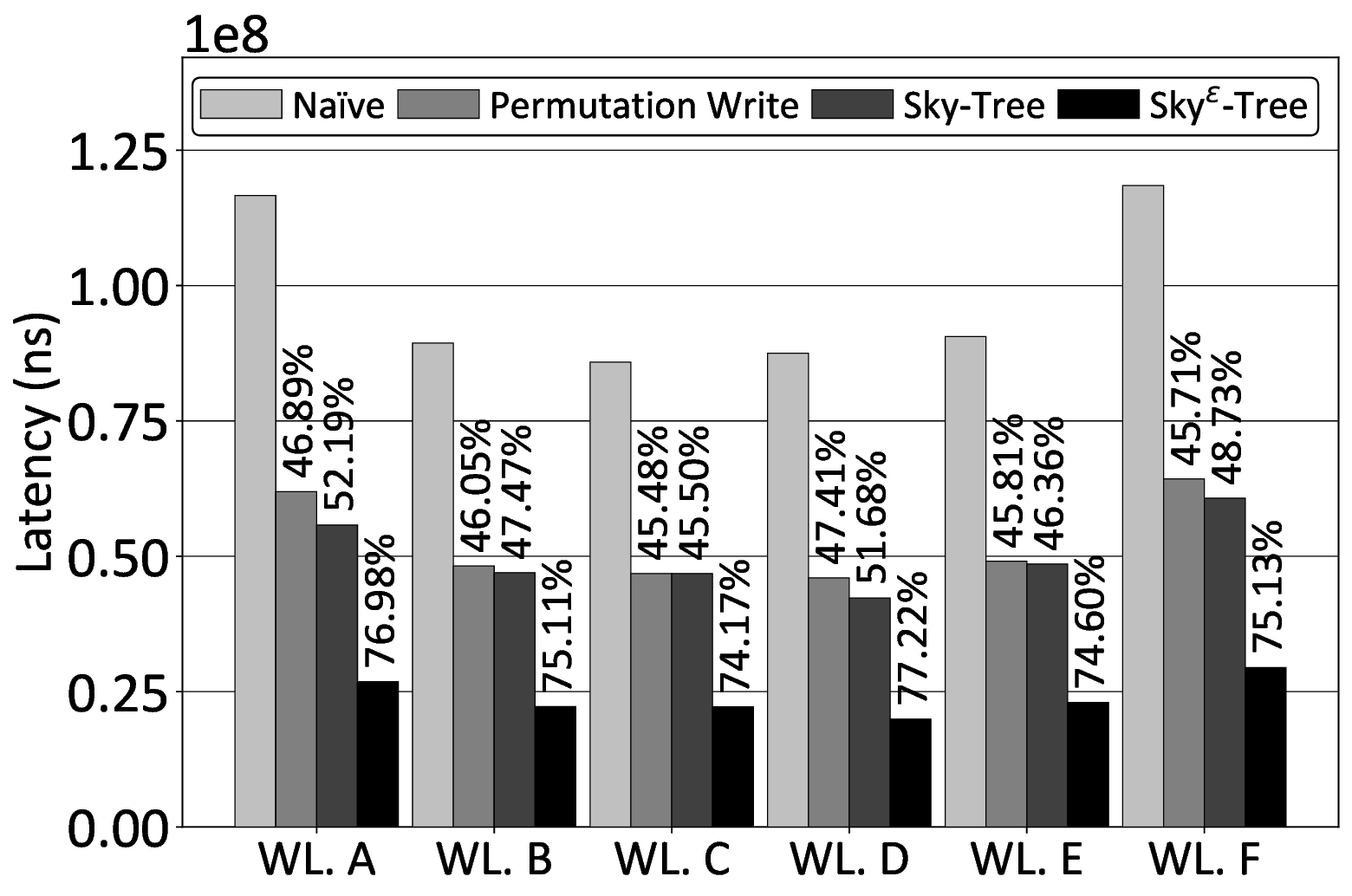}
		\vspace{-0.15in}
		\caption{Latency comparison with 10,000 entries and 8 B words.}
		\label{fig:l}
	\end{minipage}\hspace{0.03in}
	\begin{minipage}[h]{0.36\textwidth}
		\centering
		\includegraphics[height=1.65in] {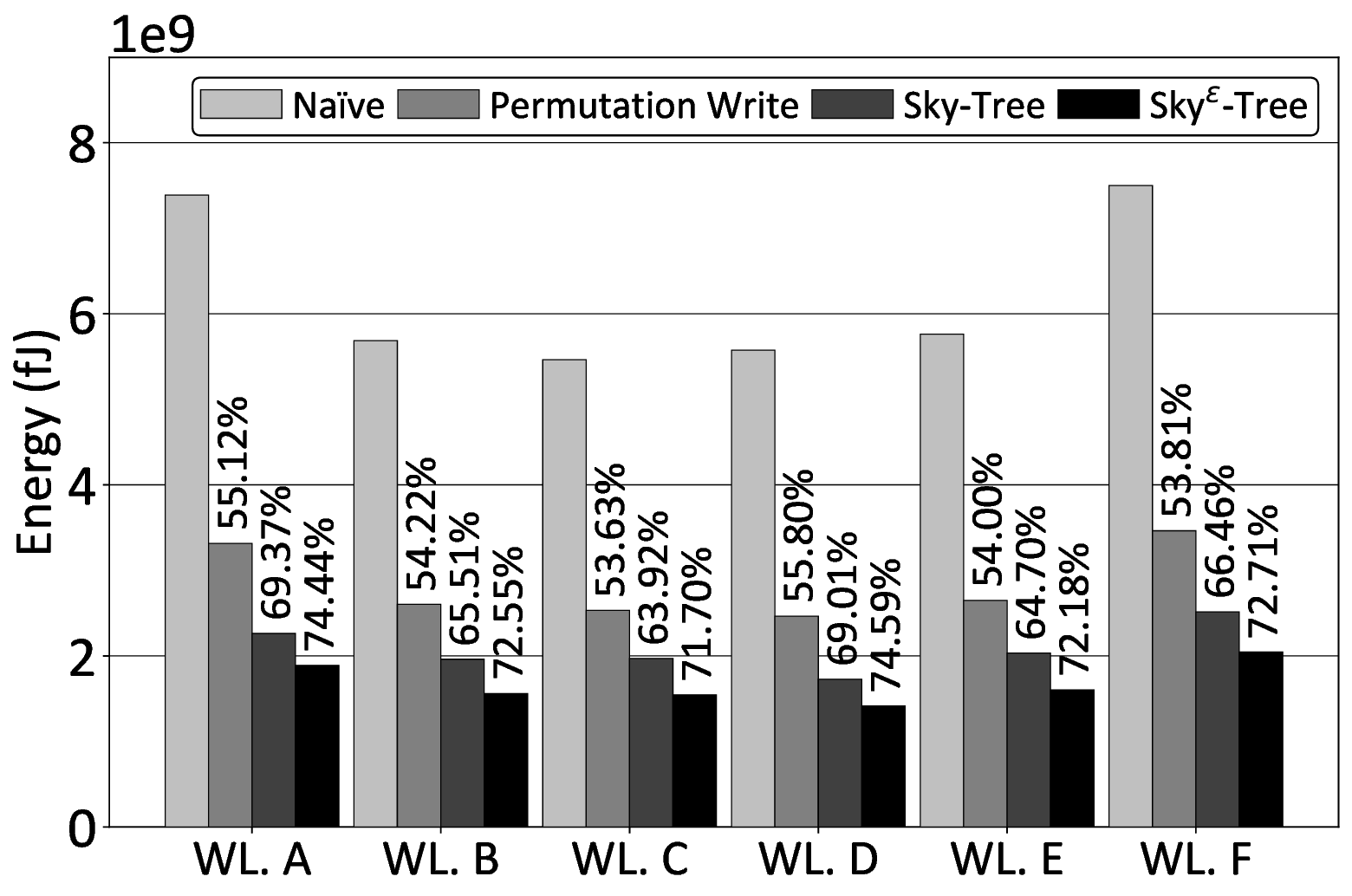}
		\vspace{-0.15in}
		\caption{Energy consumption comparison with 10,000 entries and 8 B words.}
		\label{fig:e}
	\end{minipage}\hspace{0.03in}
	\begin{minipage}[h]{0.26\textwidth}
		\centering
		\includegraphics[height=1.65in] {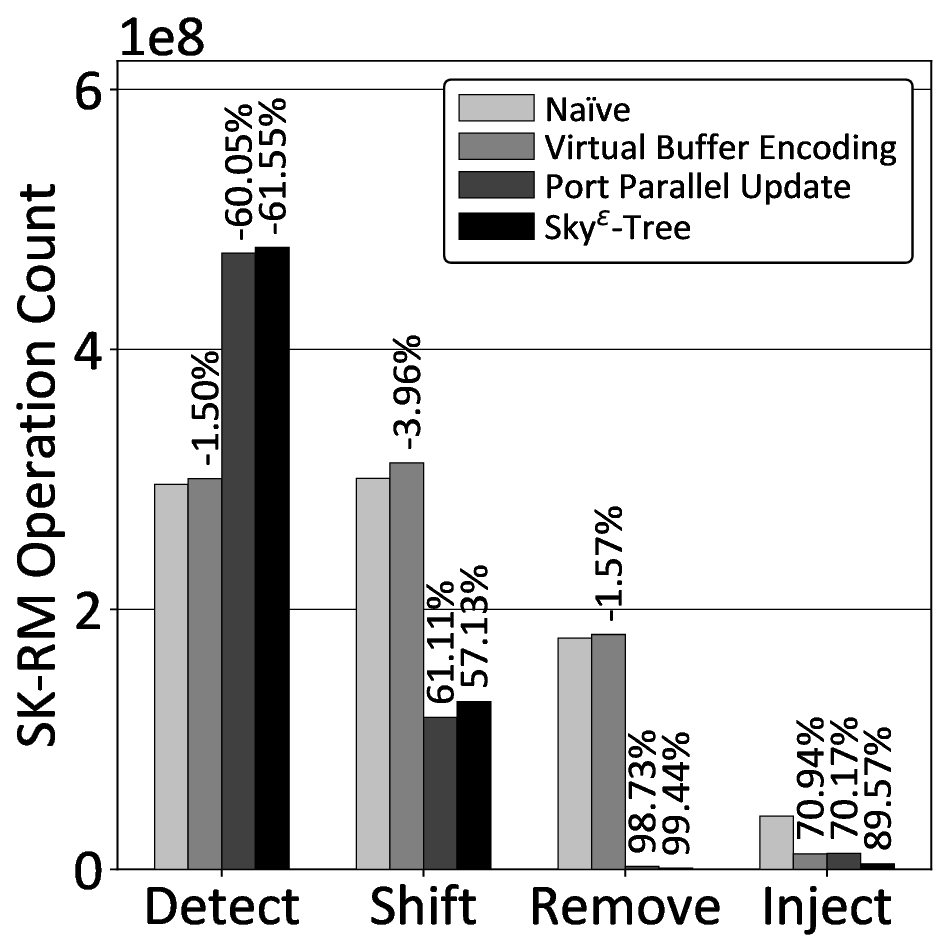}
		\vspace{-0.15in}
		\caption{Op. Cnt. comparison with 10,000 entries and 8 B words.}
		\label{fig:c}
	\end{minipage}
	\vspace{-0.15in}
\end{figure*}

\begin{figure*}[t]
	\centering
	\begin{subfigure}{0.16\textwidth}
		\centering
		\includegraphics[height=1.4in]{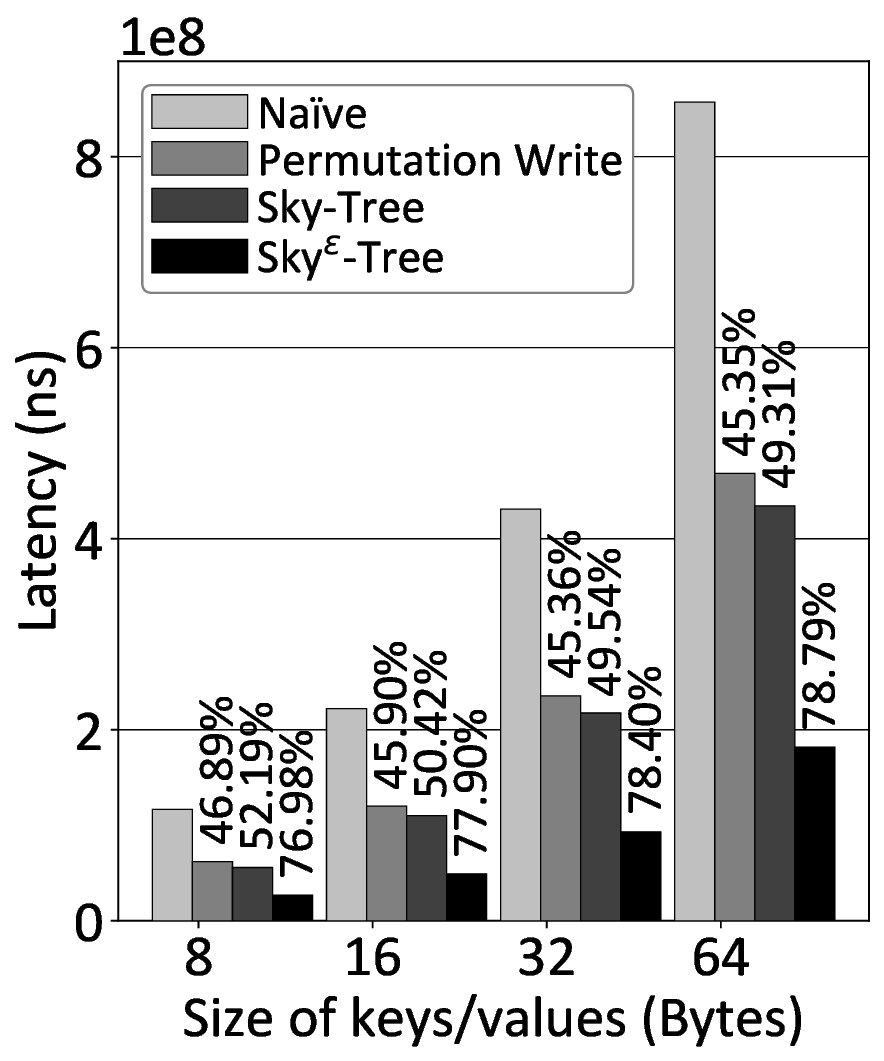}
		\vspace{-0.15in}
		\caption{Workload a} \label{fig:1a}
	\end{subfigure}\hspace{0.02in}
	\begin{subfigure}{0.16\textwidth}
		\centering
		\includegraphics[height=1.4in]{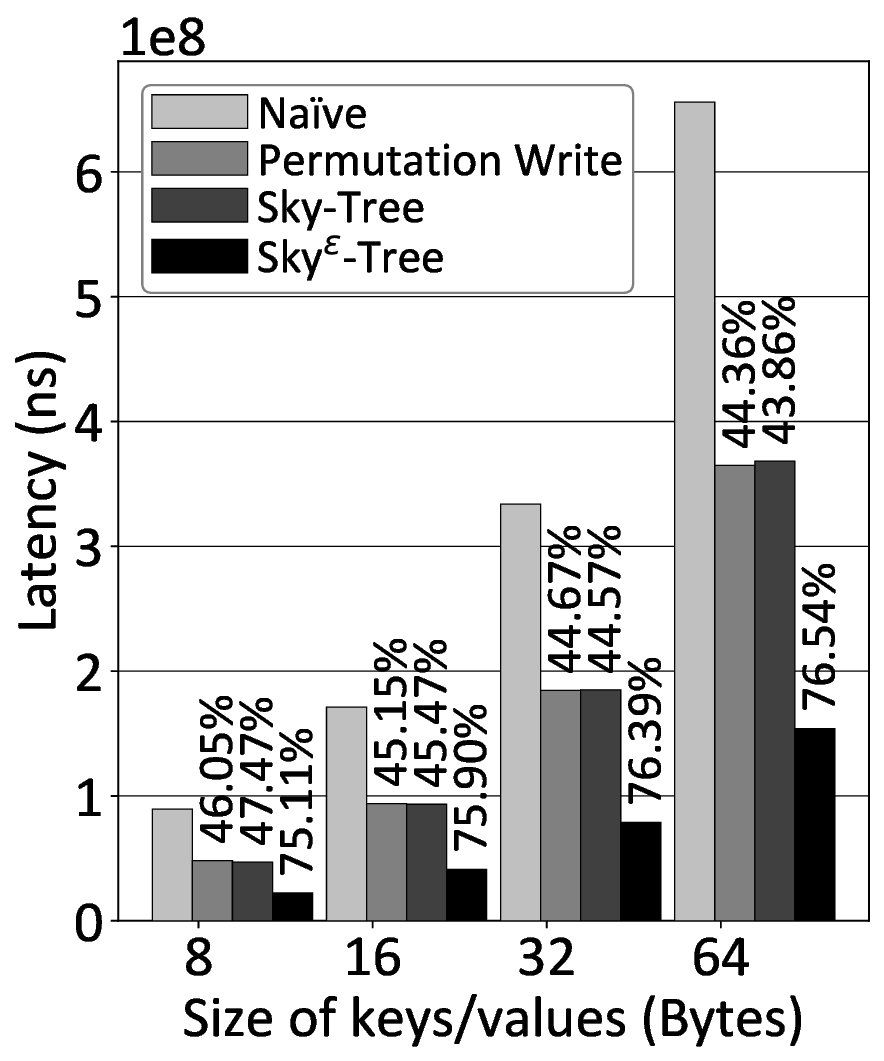}
		\vspace{-0.15in}
		\caption{Workload b} \label{fig:1a}
	\end{subfigure}\hspace{0.02in}
	\begin{subfigure}{0.16\textwidth}
		\centering
		\includegraphics[height=1.4in]{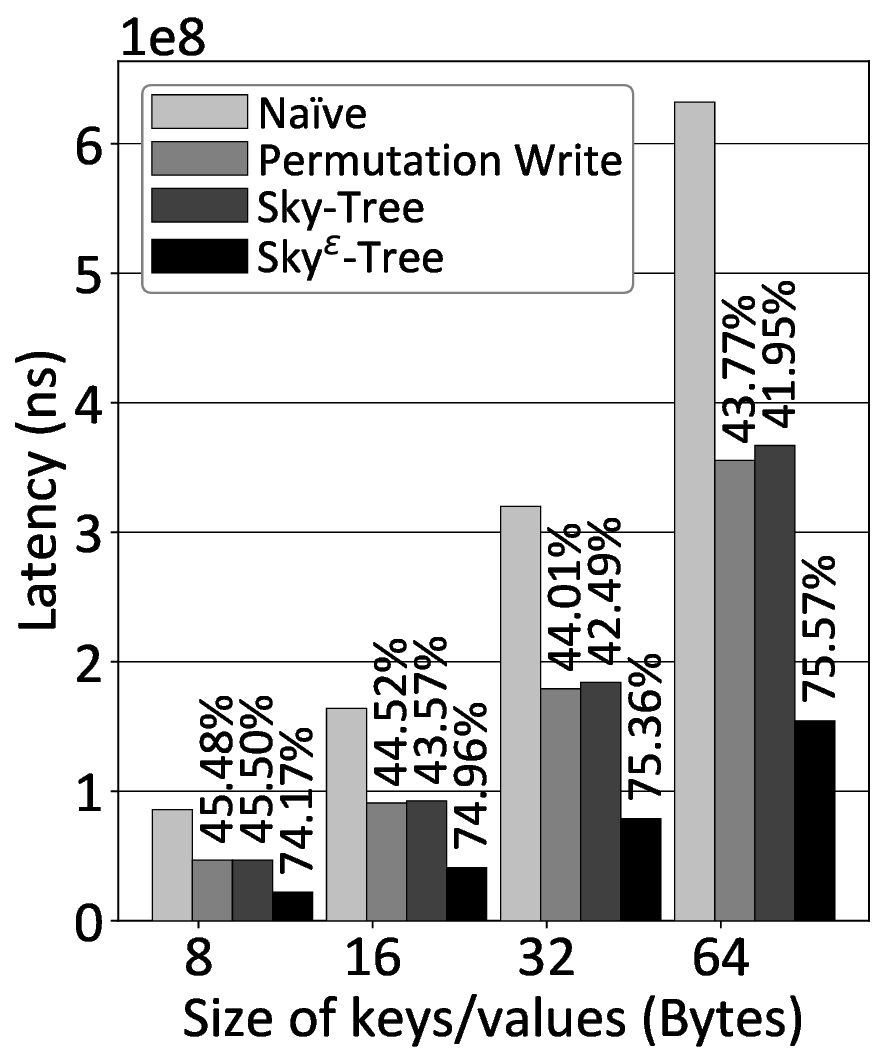}
		\vspace{-0.15in}
		\caption{Workload c} \label{fig:1a}
	\end{subfigure}\hspace{0.02in}
	\begin{subfigure}{0.16\textwidth}
		\centering
		\includegraphics[height=1.4in]{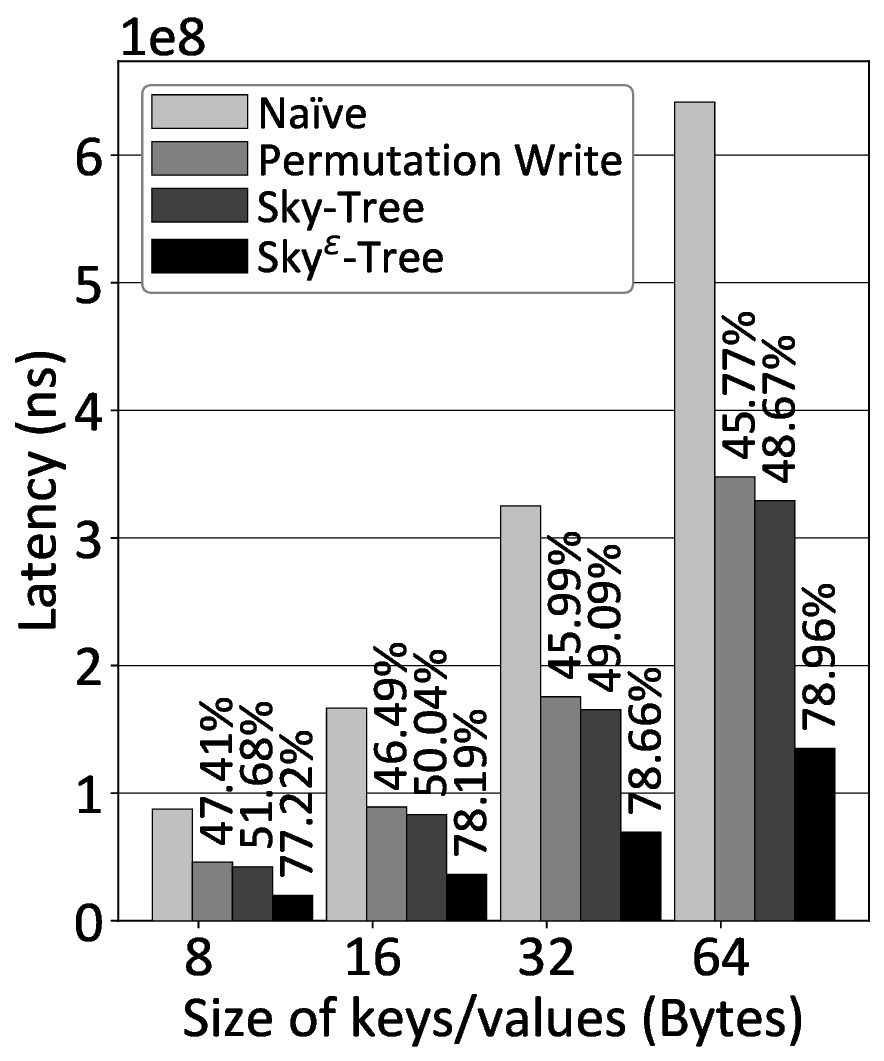}
		\vspace{-0.15in}
		\caption{Workload d} \label{fig:1a}
	\end{subfigure}\hspace{0.02in}
	\begin{subfigure}{0.16\textwidth}
		\centering
		\includegraphics[height=1.4in]{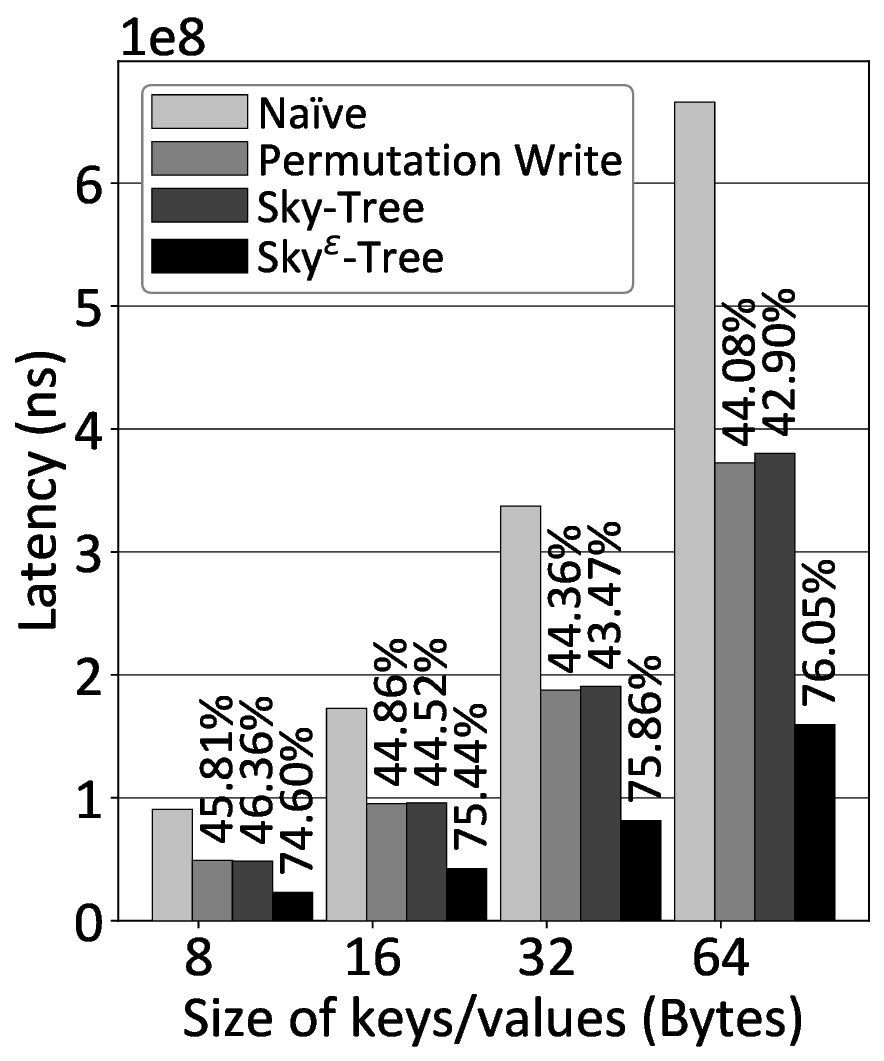}
		\vspace{-0.15in}
		\caption{Workload e} \label{fig:1a}
	\end{subfigure}\hspace{0.02in}
	\begin{subfigure}{0.16\textwidth}
		\centering
		\includegraphics[height=1.4in]{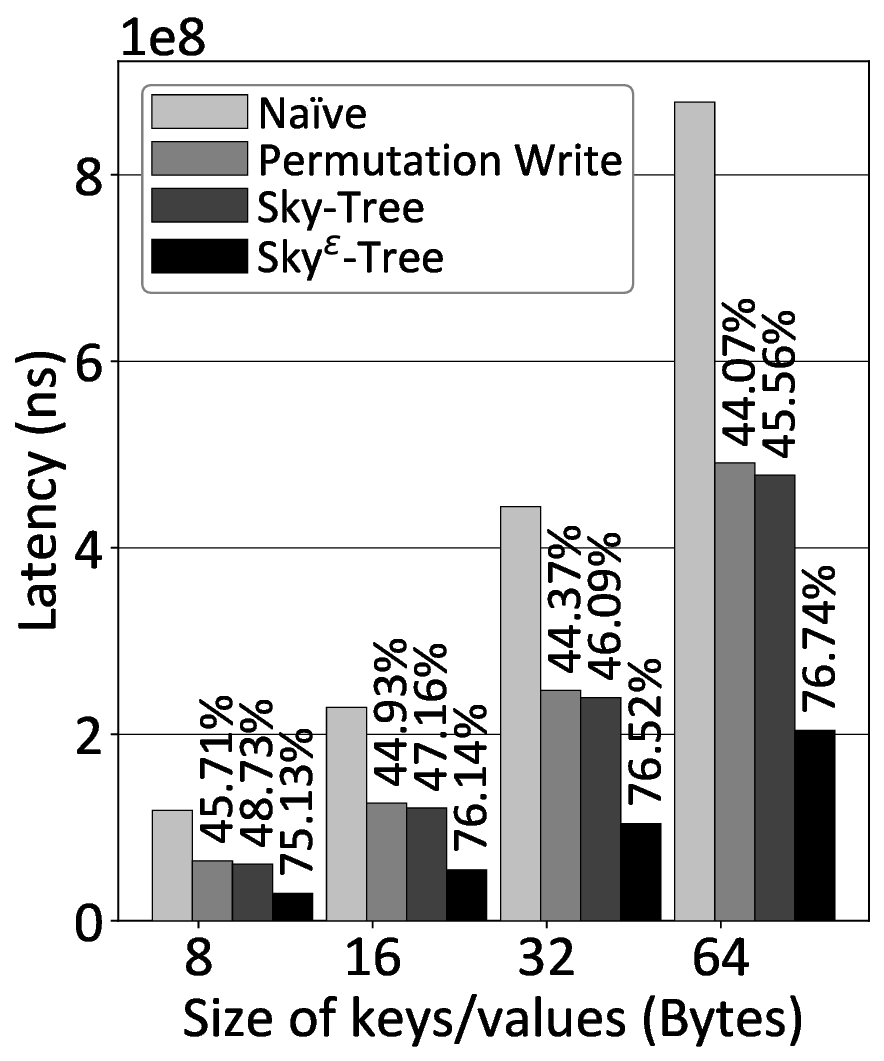}
		\vspace{-0.15in}
		\caption{Workload f} \label{fig:1a}
	\end{subfigure}%
	\vspace{-0.1in}
	\caption{Latency comparison for each workload with different sizes of keys/values (i.e., words) and 10,000 entries.}
	\label{fig:diffsize}
	\vspace{-0.15in}
\end{figure*}

\begin{figure*}[t]
	\centering
	\begin{subfigure}{0.16\textwidth}
		\includegraphics[height=1.4in]{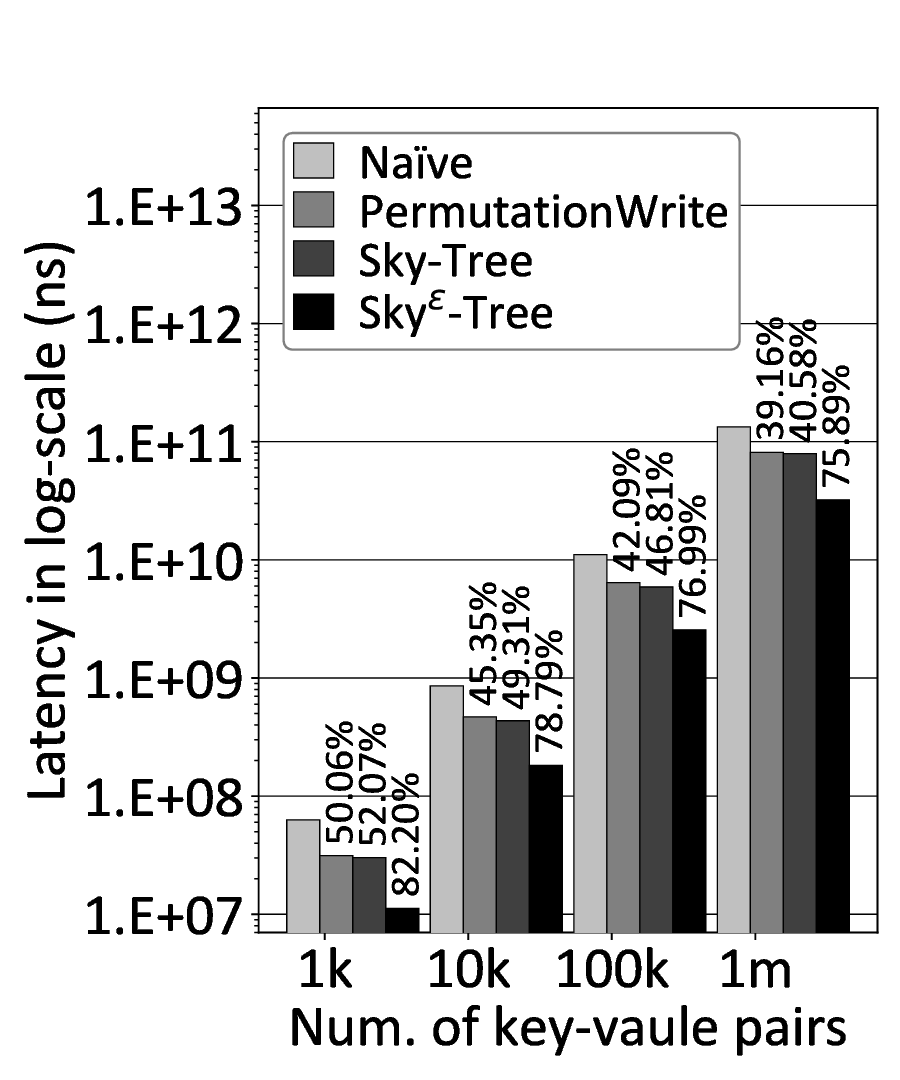}
		\vspace{-0.15in}
		\caption{Workload a} \label{fig:1a}
	\end{subfigure}\hspace{0.01in}
	\begin{subfigure}{0.16\textwidth}
		\includegraphics[height=1.4in]{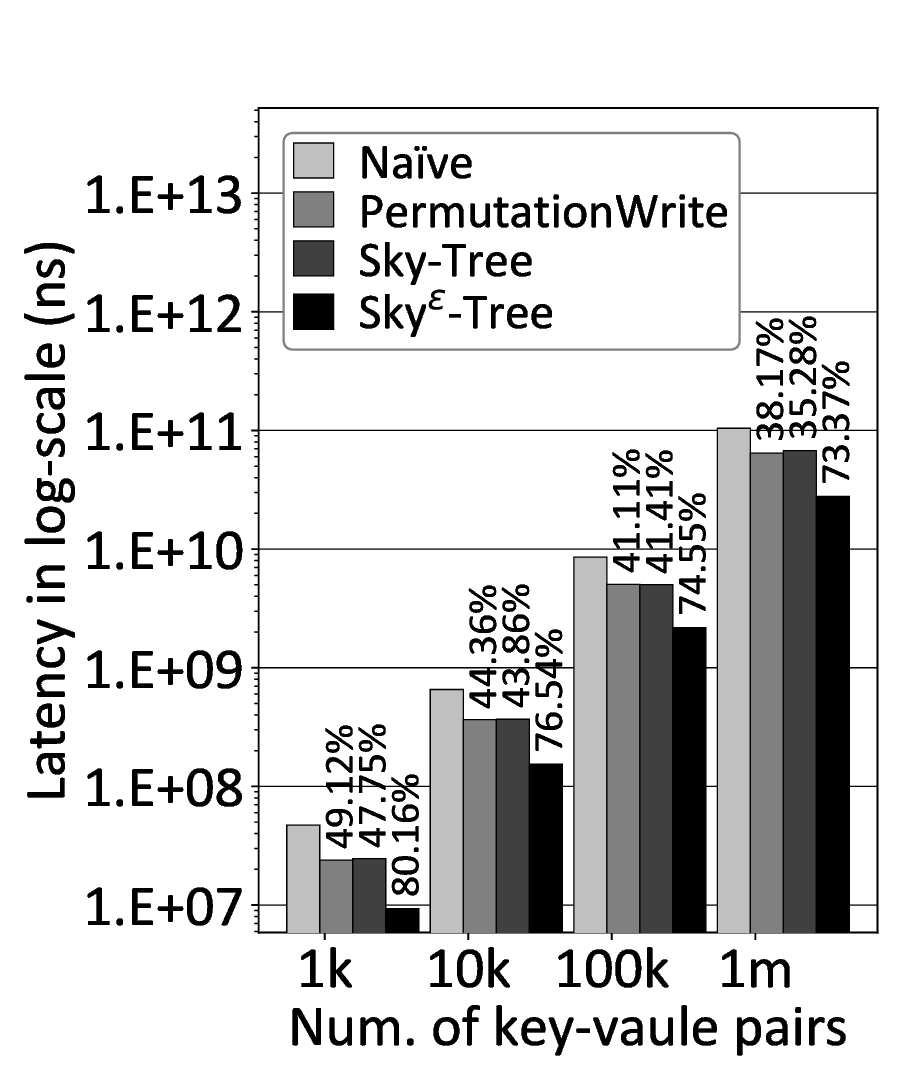}
		\vspace{-0.15in}
		\caption{Workload b} \label{fig:1a}
	\end{subfigure}\hspace{0.01in}
	\begin{subfigure}{0.16\textwidth}
		\includegraphics[height=1.4in]{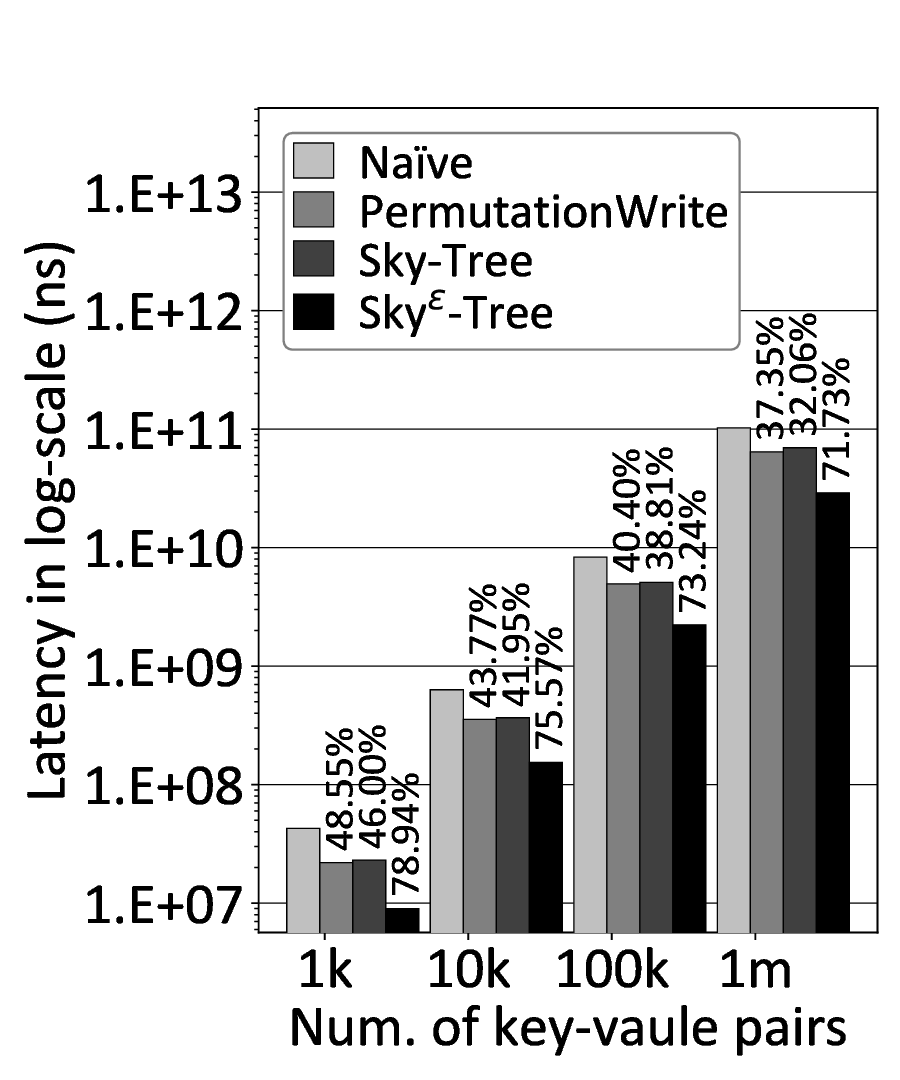}
		\vspace{-0.15in}
		\caption{Workload c} \label{fig:1a}
	\end{subfigure}\hspace{0.01in}
	\begin{subfigure}{0.16\textwidth}
		\includegraphics[height=1.4in]{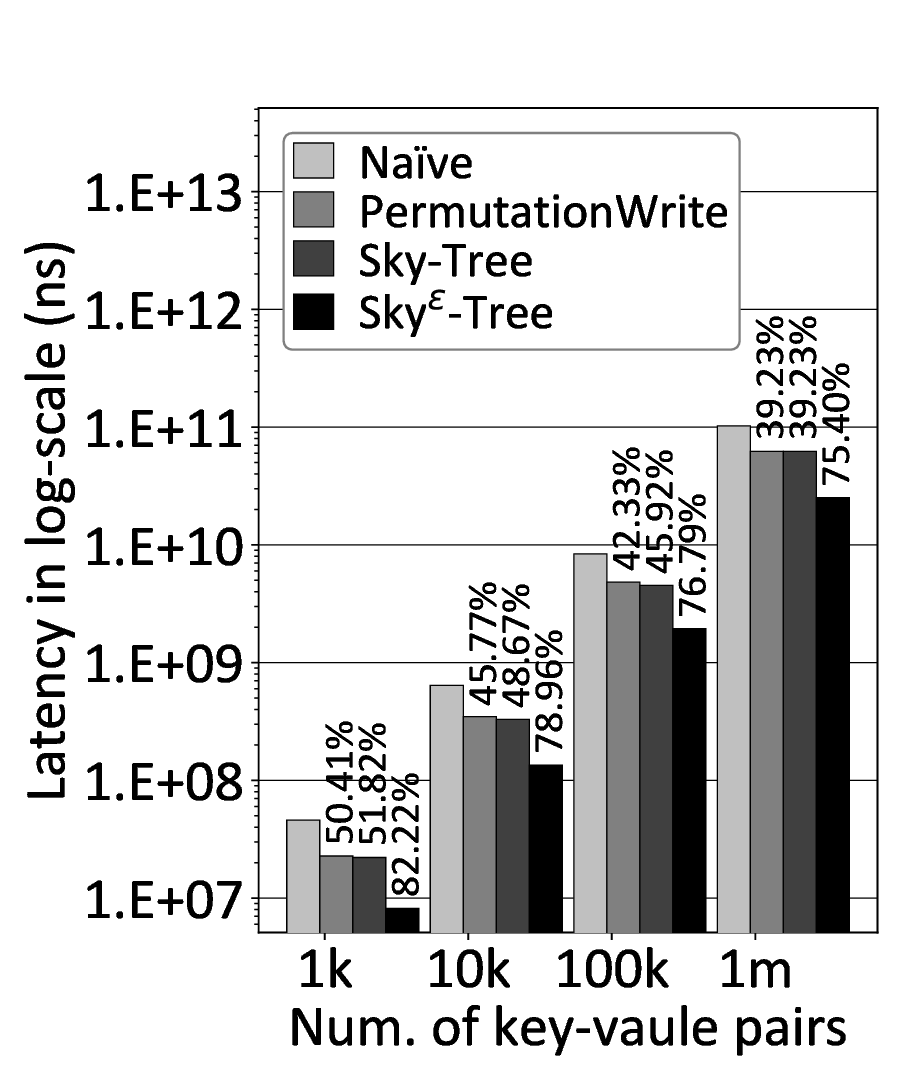}
		\vspace{-0.15in}
		\caption{Workload d} \label{fig:1a}
	\end{subfigure}\hspace{0.01in}
	\begin{subfigure}{0.16\textwidth}
		\includegraphics[height=1.4in]{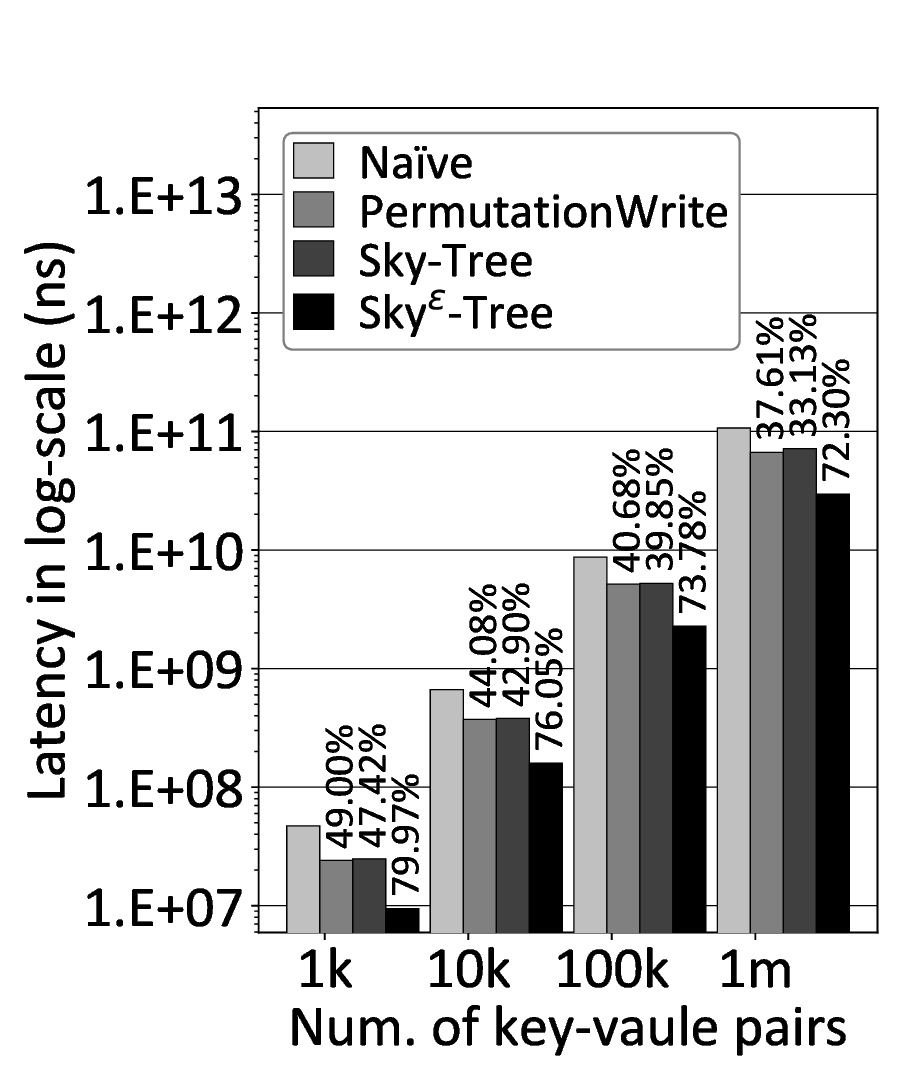}
		\vspace{-0.15in}
		\caption{Workload e} \label{fig:1a}
	\end{subfigure}\hspace{0.01in}
	\begin{subfigure}{0.16\textwidth}
		\includegraphics[height=1.4in]{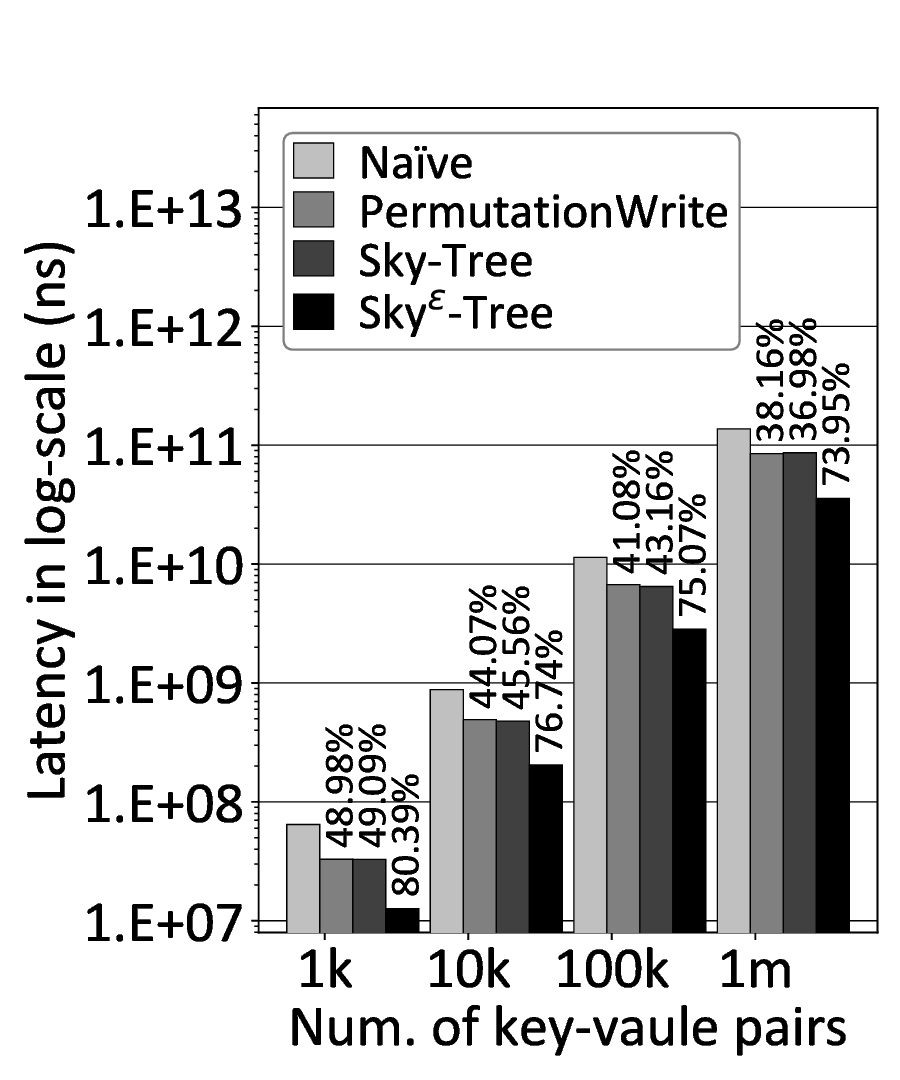}
		\vspace{-0.15in}
		\caption{Workload f} \label{fig:1a}
	\end{subfigure}%
	\vspace{-0.15in}
	\caption{Latency comparison for each workload with different numbers of key-value pairs and 8 B words.}
	\label{fig:diffnum}
\end{figure*}

\balance
\section{Performance Evaluation} \label{S:evaluation}
To evaluate the effectiveness of the proposed Sky$^{\varepsilon}$-tree, in this section, we present the evaluation results, including the latency and energy consumption comparisons, based on extensive simulations. Notably, both the word-based and bit-interleaved mapping architectures are investigated.

\subsection{Experiment Setup}
\setlength{\emergencystretch}{3em}
Based on an in-house SK-RM simulator and RTSim~\cite{khan2019rtsim}, two sets of experiments were conducted to simulate the behaviors of Sky$^{\varepsilon}$-tree on word-based and bit-interleaved mapping architectures, respectively. 
Originally, RTSim was developed as an architectural-level simulation framework built on top of NVMain for Domain Wall Memories (DWM), where trace files with memory instructions are used to perform simulations. 
For our study, a modification was made to support SK-RM operations together with metrics to keep track of the number of skyrmions that are created or destroyed in the duration of the run. 
Since RTSim assumes a bit-interleaved data organization with limited bit fields for every input trace pattern, we additionally developed an in-house SK-RM simulator to take word-based mapping with different key-value sizes into account and calculate the operation counts at bit-level. 

The workload was generated via Yahoo! Cloud Serving Benchmark~\cite{10.1145/1807128.1807152}, which provides key-value pairs according to predefined workload patterns (i.e., workload a to f), and their characteristics are summarized in Table~\ref{wl}. Please note that even though workload c consists of data reads only, data writes are still induced during the initialization of YCSB. The energy and latency parameters of the considered SK-RM are given in Table~\ref{skyr_laten_opera}, which were added to the configuration file of RTSim and considered in our in-house simulator as well.
In the following experiments, the size of a word (i.e., $WORD\_SIZE$), the number of key-value pairs, and the number of access ports are varied as knobs for understanding the performance of the proposed Sky$^{\varepsilon}$-tree under different scenarios.

\begin{table}[h]
	\centering
	\footnotesize
	\caption{Characteristics of YCSB Workloads a to f.\vspace{-0.03in}}
	\label{wl}
	\begin{tabular}{lcccc}
		\toprule
		Workload & Read & Update & Insert & Distribution \\
		\midrule
		a & 50\% & 50\% &  & zipfian \\
		\midrule
		b & 95\% &  & 5\% & zipfian \\
		\midrule
		c & 100\% &  &  & zipfian \\
		\midrule
		d & 5\% & 95\% &  & latest \\
		\midrule
		e & 95\% &  & 5\% & zipfian \\
		\midrule
		f & 50\% & 50\% &  & zipfian \\
		\bottomrule
	\end{tabular}
\end{table}
\begin{table}[h]
	\centering
	\footnotesize
	\caption{Energy and latency parameters of SK-RM~\cite{LLC}.\vspace{-0.03in}}
	\label{skyr_laten_opera}
	\begin{tabular}{lcccc}
		\toprule
		Operations & Detect & Shift & Remove & Inject \\
		\midrule
		Energy & 2~fJ & 20~fJ & 20~fJ & 200~fJ \\
		\midrule
		Latency & 0.1~ns & 0.5~ns & 0.8~ns & 1~ns \\
		\bottomrule
	\end{tabular}
\end{table}

For the word-based mapping architecture, the experimental comparisons included \textit{na\"ive}, \textit{PW} and \textit{Sky-tree}. While \textit{na\"ive} refers to the remove-all and inject-all update method, \textit{PW} and \textit{Sky-tree} refer to the permutation-write design~\cite{9218642} and the B$^{+}$-trees-based Sky-tree design~\cite{9925692}, respectively. In this experiment, the concept of Sky-tree was adapted to B$^{\varepsilon}$-trees and the implemented methods included the bit-level binary search method for facilitating the key comparison during queries, the node-based skyrmion recycler to repurpose unused words at the end of racetrack as recycle sites for residual skyrmions left after data updates, and the track-based splitting operations for splitting internal nodes without key rewriting. For the bit-interleaved mapping architecture, Sky$^{\varepsilon}$-tree was compared with \textit{na\"ive} and \textit{DCW} methods, where \textit{na\"ive} refers to the remove-all and inject-all update method, and \textit{DCW} refers to the data-comparison strategy~\cite{LLC}. Notably, within RTSim the shift method was configured as lazy for all experiments.


\subsection{Results -- Word-based Mapping}
Figures~\ref{fig:l} and~\ref{fig:e} first show the latency and energy consumption comparisons. Note that the percentages are calculated by comparing with the \textit{na\"ive} data update method for the word-based mapping. Accordingly, when compared with the \textit{na\"ive} data update method, the Sky$^{\varepsilon}$-tree can effectively reduce the latency and energy consumption by up to 77.22\% and 74.59\%, respectively. When compared with \textit{PW}, the average latency improvements and energy reduction are 50.02\% and 40.83\%, respectively. These improvements mainly come from the proposed port parallel update that reduces the number of shift operations and the virtual buffer encoding that lowers the number of to-be-written bits during flush or split operations. When compared with the previous Sky-tree design, the improvements are 52.34\% and 19.39\% on average for latency and energy. In other words, directly deploying the methods of Sky-tree onto B$^{\varepsilon}$-trees does not yield the best performance and energy improvements.

To verify the observations that improvements are from the reduced number of shift and inject operations, the breakdown of each operation count is shown in Figures~\ref{fig:c}, and the individual method of the Sky$^{\varepsilon}$-tree is solely enabled for clearness. Accordingly, it can be observed the number of shift operations drops after the port parallel update is enabled. Similarly, it can be observed that the number of inject operations is improved by 70.94\% with the virtual buffer encoding enabled. In summary, the proposed Sky$^{\varepsilon}$-tree can effectively reduce the skyrmion operations for performance and energy efficiency improvement. Next, to further understand how the word-based Sky$^{\varepsilon}$-tree performs with different sizes of keys/values and different sizes of datasets, the results are summarized in Figure~\ref{fig:diffsize} and Figure~\ref{fig:diffnum} for each workload. 

Figure~\ref{fig:diffsize} shows that the latency reduction of the proposed Sky$^{\varepsilon}$-tree grows when the size of keys/values becomes larger. This growing trend can be attributed to the proposed virtual buffer encoding -- when the size of values becomes larger, the difference in the number of bits between the indices and the separated values also becomes larger. Therefore, the virtual buffer encoding can reduce the inject operations during flush/split. On the other hand, as Figure~\ref{fig:diffnum} shows, the latency improvement drops slightly when the size of datasets grows toward 1 million key-value pairs. We observe that this slight performance drop is a result of a growing number of detect operations when the size of datasets becomes larger. In other words, the portion of detect latency becomes larger in the total latency and, in turn, compresses the benefit of the reduced number of inject and shift operations. Nonetheless, the proposed Sky$^{\varepsilon}$-tree still achieves up to $75.89\%$ latency reduction with the 1-million dataset.

\begin{figure*}
	\centering
	\begin{minipage}[h]{0.37\textwidth}
		\centering
		\includegraphics[height=1.69in] {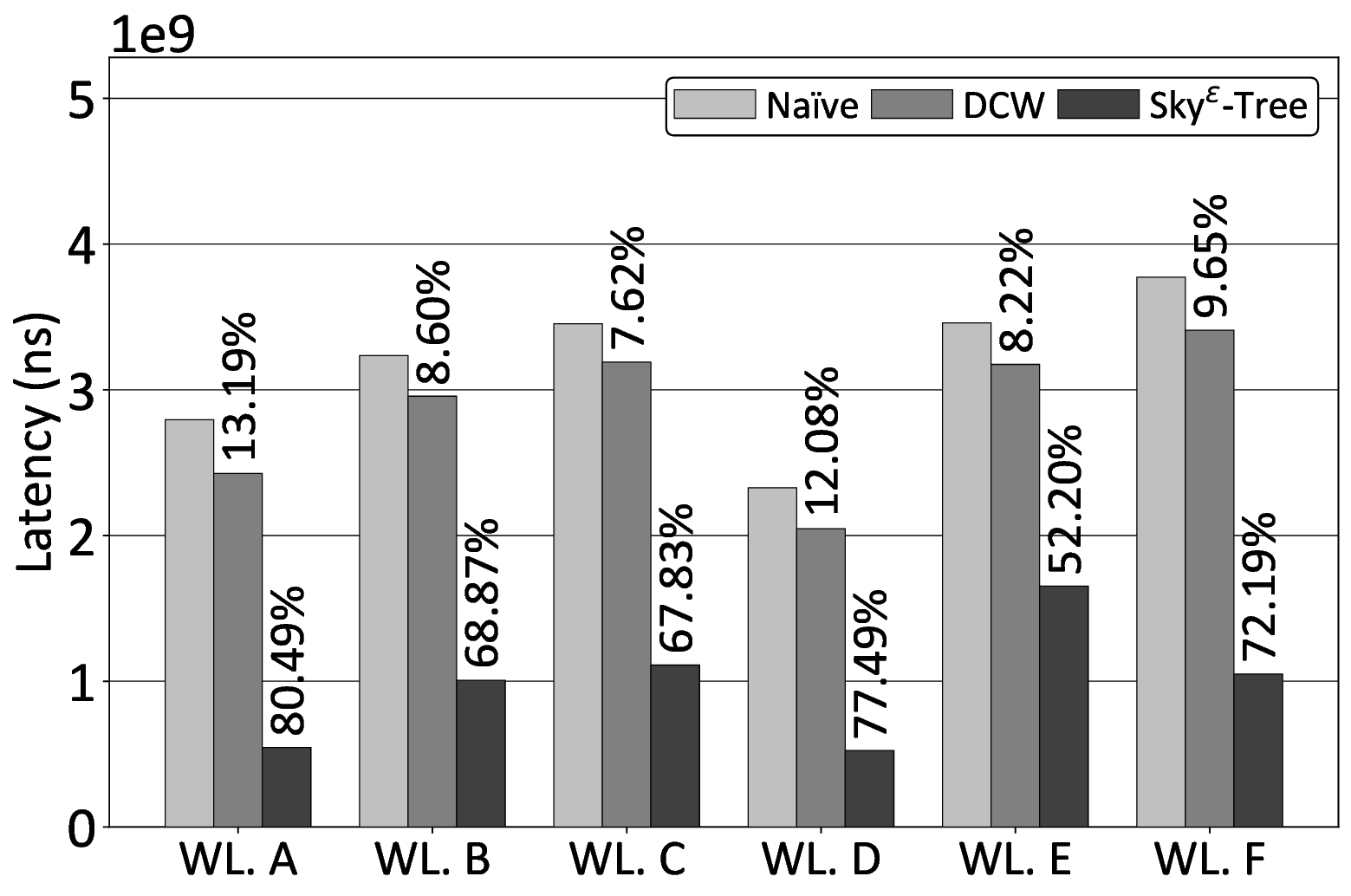}
		\vspace{-0.15in}
		\caption{Bit-interleaved: Latency comparison with 10,000 entries and 8 B word.}
		\label{fig:bl}
	\end{minipage}\hspace{0.03in}
	\begin{minipage}[h]{0.37\textwidth}
		\centering
		\includegraphics[height=1.69in] {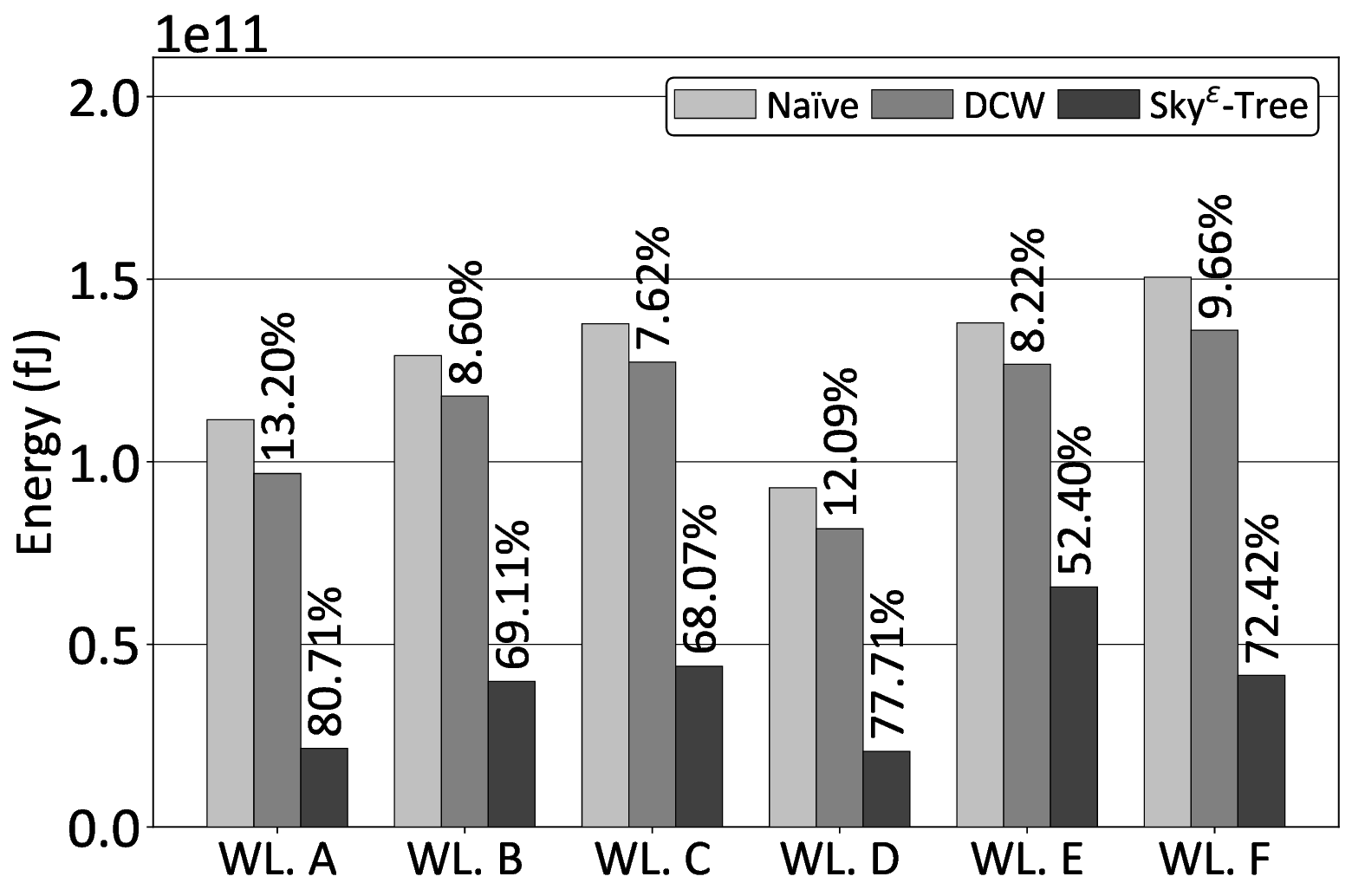}
		\vspace{-0.15in}
		\caption{Bit-interleaved: Energy cons. comparison with 10,000 entries and 8 B word.}
		\label{fig:be}
	\end{minipage}\hspace{0.03in}
	\begin{minipage}[h]{0.23\textwidth}
		\centering
		\includegraphics[height=1.69in] {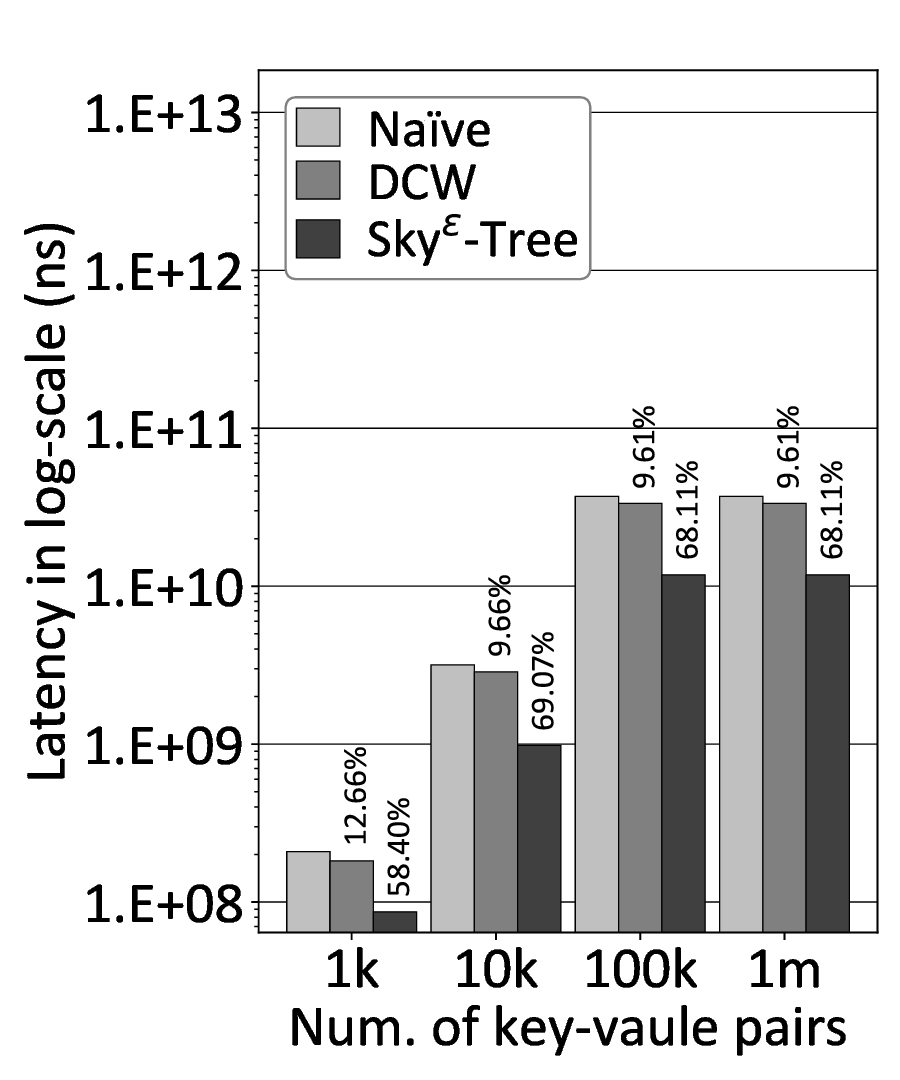}
		\vspace{-0.15in}
		\caption{Bit-interleaved: lat. cf. with diff. nums. of pairs.}
		\label{fig:b}
	\end{minipage}
\end{figure*}

\vspace{0.05in}
\subsection{Results -- Bit-interleaved Mapping}
\vspace{0.05in}
The latency and energy comparisons are shown in Figures~\ref{fig:bl} and~\ref{fig:be}. First, we observe that adopting bit-interleaved mapping through the proposed data layout (See Section~\ref{sub:bim}) is effective and can achieve up to 80.49\% and 80.71\% of latency and energy reductions. Similar to the observations for the word-based mapping architecture, these improvements mainly come from the reduced amount of \texttt{shift} and \texttt{inject} operations. The difference is that the parallel port update of word-based mapping requires shifting words out and back across the access ports; the bit-interleaved version only needs to align to-be-updated words with access ports and perform bit-comparison writes directly. In other words, the number of shift operations could be even lower with the bit-interleaved mapping. 

On the other hand, the size of the dataset is also varied for the bit-interleaved Sky$^{\varepsilon}$-tree; the results are shown in Figure~\ref{fig:b}. Different from the slight performance drop observed in the case of word-based mapping, the improvement slightly grows with larger datasets. The rationale behind is that the \texttt{detect} operations on the bit-interleaved mapping architecture can be triggered simultaneously. 
Hence, the accumulated of \text{detect} operations is greatly lower while the benefit of the proposed Sky$^{\varepsilon}$-tree still persists. Last but not the least, the improvement actually grows from 1,000 to 10,000 datasets, because the virtual buffer encoding is more effective with larger datasets, which induce numerous flush/split operations.

\section{Conclusion} \label{S:conclusion}
In this work, we proposed Sky$^{\varepsilon}$-tree to enhance the efficiency of B$^{\varepsilon}$-trees deployed on Skyrmion Racetrack Memory (SK-RM). By leveraging the unique characteristics of SK-RM, Sky$^{\varepsilon}$-tree optimizes data layout and update methods. The proposed virtual buffer encoding separates keys and values in buffer entries for reducing the number of skyrmion injections. The port parallel update enables simultaneous writing of multiple key-value pairs to reduce the number of shift operations required during batched updates. 
The experimental results show that the proposed methods, virtual buffer encoding and parallel port update, are effective on both word-based and bit-interleaved mapping architectures. In summary, by focusing on minimizing skyrmion injections and optimizing data handling, Sky$^{\varepsilon}$-tree strikes a balance of performance and storage density while retaining the essential functionalities of B$^{\varepsilon}$-trees.


\thispagestyle{empty}

\bibliographystyle{ACM-Reference-Format}
\bibliography{sample}    

\end{document}